\documentclass[pre,aps,groupaddress,showpacs,prd,onecolumn]{revtex4-1}
\usepackage{epsfig,amsmath,graphicx,amssymb,overpic}

\usepackage{graphicx}
\usepackage{dcolumn}
\usepackage{bm}
\usepackage{graphicx}
\usepackage{subfigure}

\def\be{\begin{equation}}
\def\ee{\end{equation}}
\def\bee{\begin{eqnarray}}
\def\ene{\end{eqnarray}}
\def\bes{\begin{subequations}}
\def\ees{\end{subequations}}

\begin{document}

\title{Generalized perturbation $(n, M)$-fold Darboux transformations and multi-rogue wave structures for the modified
self-steepening nonlinear Schr\"{o}dinger equation}

\author{Xiao-Yong Wen$^{1,2}$}
\email{wenxiaoyong@amss.ac.cn}
\author{Yunqing Yang$^{1,3}$}
\author{Zhenya Yan$^1$}
\email{Corresponding author, Email: zyyan@mmrc.iss.ac.cn}
\affiliation{\vspace{0.1in}  $^1$Key Laboratory of Mathematics Mechanization, Institute of Systems Science, AMSS, Chinese Academy of Sciences, Beijing 100190, China \\
$^2$Department of Mathematics, School of Applied Science, Beijing Information
    Science and Technology University, Beijing  100192, China\\
    $^3$School of Mathematics, Physics and Information Science, Zhejiang Ocean University, Zhoushan 316004, China\vspace{0.1in}}

\date{\vspace{0.1in} 31 December 2014, Phys. Rev. E  {\bf 92}, 012917 (2015)}

\begin{abstract}
\baselineskip=15pt
In this paper, a novel, simple, and constructive method is presented to find the generalized perturbation $(n, M)$-fold Darboux transformations (DTs) of the modified nonlinear Schr\"odinger (MNLS) equation in terms of fractional forms of determinants. In particular, we apply the generalized perturbation $(1, N-1)$-fold DTs to find its explicit multi-rogue wave solutions. The wave structures of these rogue wave solutions of the MNLS equation are discussed in detail for different parameters, which display abundant interesting wave structures including the triangle and pentagon, etc. and may be useful to study the physical mechanism of multi-rogue waves in optics. The dynamical behaviors of these multi-rogue wave solutions are illustrated using numerical simulations. The same Darboux matrix can also be used to investigate the Gerjikov-Ivanov equation such that its multi-rogue wave solutions and their wave structures are also found.  The method can also be extended to find multi-rogue wave solutions of other nonlinear integrable equations.
\end{abstract}


\maketitle
\baselineskip=15pt

\section{Introduction}

In general, it is difficult to directly seek for exact solutions including solitons of nonlinear wave equations in the soliton theory and integrable system. If one can find a proper Lax pair~\cite{lax} of a nonlinear wave equation, then, based on its Lax pair, one may find its solutions by means of some transformations such as the inverse scattering transformation~\cite{ivs, ivs2,ivs3},
the Riemann-Hilbert scattering method~\cite{RH}, the Darboux transformation~\cite{dt,dt2,dt3}, and so on. The Darboux transformation~\cite{dt,dt2,dt3} is regarded as a powerful approach to study solutions of nonlinear integrable systems in terms of their corresponding Lax pairs, which originally arising from Darboux's study~\cite{dt82} of the linear Sturm-Liouville equation, $\psi_{xx}+[\lambda-u(x)]\psi=0$, which is also called the linear Schr\"odinger equation with the external potential in quantum mechanism. Up to now, the Darboux transformation has been used to investigate many types of solutions of nonlinear integrable equations including multi-soliton solutions, breathers, periodic solutions, and rational solutions. In fact, the Darboux transformation exhibits the different forms in the studying nonlinear integrable sysytems, in which the following three main types of Darboux transformations were usually used: Type i) the Darboux transformation without the Darboux matrix (see, e.g., Ref.~\cite{nail88}); Type ii) the Darboux transformation with the iteration Darboux matrix (see , e.g., Ref.~\cite{dt}); Type iii) the Darboux transformation with the $N$-order Darboux matrix whose elements being the polynomials of the spectral parameters (see, e.g., Refs.~\cite{dt2,dt3}).

Recently, the modified Darboux transformation related to Type i) was used to find multi-rogue wave solutions of the self-focusing nonlinear Sch\"odinger (NLS) equation from the initial plane wave solution~\cite{nail09}. Moreover, the generalized Darboux transformation related to Type ii) with the initial plane wave solution was also used to investigate multi-rogue wave solutions of the self-focusing NLS equation~\cite{guo1}. In fact, there exist some other methods to study rogue wave solutions of many nonlinear wave equations with constant or varying coefficients (see, e.g.,~\cite{rw1,yang2,yan,rw2,lak}). It is still an important topic to present more powerful and simple methods to find other types of solutions including multi-rogue wave solutions of nonlinear integrable equations.

 To our best knowledge, the Type iii) of Darboux transformations and their some extensions were not used to investigate other types of solutions (e.g., multi-rogue wave solutions) of integrable equations except for their multi-soliton-type solutions. In this paper, we will present a novel approach to construct the generalized perturbation $(n, M)$-fold Darboux transformation of  nonlinear integrable equations based on their Type iii) of  Darboux transformations and the Darboux matrix in terms of the Taylor series expansion for the parameter and a limit procedure such that their multi-rogue wave solutions can be obtained.

The modified nonlinear Schr\"{o}dinger (MNLS) equation  with the self-steepening term is given by in the form
\begin{eqnarray}
 i q_{t}+q_{xx}+i(|q|^2q)_x+2 \rho|q|^2q =0, \label{mnls}
\end{eqnarray}
where $q\equiv q(x,t)$ is the slowly varying complex envelope of the wave, $\rho$ is a real constant and $i^2=-1$, the subscript denotes the partial derivative with respect to the variables $x,\,t$, the term $i(|q|^2q)_x$ is called the self-steepening term, which causes an optical pulse to become asymmetric and steepen up at the trailing edge~\cite{ag,yang}. When $\rho=0$, Eq.(\ref{mnls}) reduces to the Kaup-Newell type of the derivative NLS equation~\cite{kn}. In fact, Eq.(\ref{mnls}) can also reduce to the derivative NLS equation via a gauge transformation~\cite{gt}.  Eq.~(\ref{mnls}) describes the short pulses propagate in a long optical fiber charaterized by a nonlinear refractive index $n(\omega, \vec{E})=n(\omega)+n_2|\vec{E}|^2$~\cite{mnls1}. Eq.~(\ref{mnls}) can  also be used to describe Alfv\'en waves propagating along the magnetic field in cold plasmas~\cite{cp} and the deep-water gravity waves~\cite{wm}. Eq.~(\ref{mnls}) was also called the perturbation NLS equation~\cite{pnls}.  Eq.~(\ref{mnls}) was shown to be completely integrable by the inverse scattering transformation~\cite{wadati}. The dynamics of optical solitons with initial phase modulation was studied~\cite{kon88}. The projection matrix method  combining the Daroux transformation and the Riemann-Hilbert problem was used to derive the DT and B\"{a}cklund transformation for Eq.~(\ref{mnls}) such that a new solution was obtained~\cite{xiao}. Its $N$-soliton solutions were also obtained in terms of the generalized Zakharov-Shabat equation ~\cite{huang} and the bilinear method~\cite{ns}. Moreover, multiple pole solutions of Eq.~(\ref{mnls}) were also found~\cite{liu}. Eq.~(\ref{mnls}) and its higher-dimensional case were also studied analytically or numerically~\cite{ha,ha2,ha3,ha4}. Recently, some rogue wave solutions of the derivative NLS
equation have been investigated~\cite{guo}.

In this paper, we will further investigate multi-rogue wave solutions of Eq.~(\ref{mnls}) via the generalized perturbation $(n ,M)$-fold Darboux transformation technique, which is a novel, simple, and constructive method. Moreover, this method with the same formal Darboux matrix can also be applied to investigate multi-rogue wave solutions of the Gerjikov-Ivanov equation~\cite{fan1}
\bee\label{gi}
   iq_t+q_{xx}-iq^2q^{*}_x+\frac12|q|^4q=0.
\ene

The rest of this paper is organized as follows. In Sec.II, we will give a brief introduction of type iii) of $N$-fold DT for Eq.~(\ref{mnls}) based on the Lax pair in Ref.~\cite{xiao}. In Sec. III, we present a novel idea to derive the generalized perturbation $(1, N-1)$-fold (using one spectral parameter) and $(n, M)$-fold (using $n$ spectral parameters) Darboux transformations for the MNLS equation (\ref{mnls}) by means of the Taylor expansion and a limit procedure related to the $N$-fold Darboux matrix. In particular, the generalized perturbation $(1, N-1)$-fold Darboux transformation is used to exhibit its multi-rogue wave solutions. Moreover we analyze the wave structures and dynamical behaviors of the multi-rogue wave solutions for  differential parameters using the numerical simulations. In Sec. IV, we illustrate the multi-rogue wave solutions of the Gerjikov-Ivanov equation equation (\ref{gi}) in terms of its generalized perturbation $(1,N-1)$-fold Darboux transformation. Finally, we will address the conclusions in Sec.V.

\section{The $N$-fold Darboux transformation}

In the following we firstly give the $N$-fold Darboux transformation of Eq.~(\ref{mnls}) in order to present our main aim, i.e., its generalized perturbation $(n, M)$-fold Darboux transformation presented in Sec. III.  The modified NLS equation (\ref{mnls}) is just a zero-curvature equation $U_t-V_x+[U,\,  V]=0$ with $[U,\, V]\equiv UV-VU$ and two $2\times 2$ matrixes $U$ and $V$ satisfying the linear iso-spectral problem (Lax pair)~\cite{xiao}
\begin{eqnarray}
\varphi_x=U \varphi, \qquad U=
                 \left(
                 \begin{array}{cc}
                 - \dfrac{i}{\lambda^2}+i\rho                          & \dfrac{q}{\lambda}  \vspace{0.1in}\\
                 -\dfrac{q^{*}}{\lambda}                                        & \dfrac{i}{\lambda^2}- i\rho  \\
                 \end{array}
                 \right), \qquad\qquad\qquad\qquad\qquad\qquad\qquad\qquad\qquad\quad  \label{lax1}
\end{eqnarray}
\begin{eqnarray}
 \varphi_{t}=V \varphi, \qquad V=
                 \left(
                 \begin{array}{cc}
        -2 i\left(\dfrac{1}{\lambda^2}-\rho\right)^2+i \dfrac{|q|^2}{\lambda^2}    & 2\left(\dfrac{1}{\lambda^3}-\dfrac{\rho}{\lambda}\right)q-\dfrac{|q|^2q}{\lambda}+i \dfrac{q_{x}}{\lambda}   \vspace{0.1in} \\
        -2\left(\dfrac{1}{\lambda^3}-\dfrac{\rho}{\lambda}\right)q^{*}+\dfrac{|q|^2q^{*}}{\lambda}+i \dfrac{q^{*}_{x}}{\lambda}
                      &  2 i\left(\dfrac{1}{\lambda^2}-\rho\right)^2-i \dfrac{|q|^2}{\lambda^2}   \\
                   \end{array}
                   \right),  \label{lax2}
\end{eqnarray}
where $\varphi=\varphi(x,t)=(\phi(x,t), \psi(x,t))^{\rm T}$ is the complex eigenfunction, $\lambda\in \mathbb{C}$ is the spectral parameter, $q=q(x,t)$ denotes the complex potential and is also the solution of Eq.~(\ref{mnls}), the subscript denotes the partial derivative with respect to the variables $x,\,t$, and the star stands for the complex conjugate of the corresponding variables.

In what follows, we introduce the gauge transformation
\begin{eqnarray}
\widetilde{\varphi}=T \varphi\ , \label{gauge}
\end{eqnarray}
where $\varphi=(\phi, \psi)^{\rm T}$ satisfies the Lax pair ~(\ref{lax1}) and (\ref{lax2}), $T$ is a $2\times 2$ Darboux matrix to be determined later, $\widetilde{\varphi}=\widetilde{\varphi}(x,t)=(\widetilde{\phi}(x,t), \widetilde{\psi}(x,t))^{\rm T}$ is required to satisfy the same formal Lax pair~(\ref{lax1}) and (\ref{lax2}) with $U$ and $V$ replaced by $\widetilde{U}$ and $\widetilde{V}$, that is,
\begin{eqnarray}
\widetilde{\varphi}_x=\widetilde{U} \widetilde{\varphi},\quad
 \widetilde{U}=\left(\begin{array}{cc}
                 - \dfrac{i}{\lambda^2}+i\rho                          & \dfrac{\widetilde{q}}{\lambda}  \vspace{0.1in}\\
                 -\dfrac{\widetilde{q}^{*}}{\lambda}                                        & \dfrac{i}{\lambda^2}- i\rho  \\
                 \end{array}
                 \right),  \qquad\qquad\qquad\qquad\qquad\qquad\qquad\qquad\qquad\quad \label{kongjian}
\end{eqnarray}
\begin{eqnarray}
\widetilde{\varphi}_{t}=\widetilde{V} \widetilde{\varphi},\quad
 \widetilde{V}=\left(
                 \begin{array}{cc}
        -2 i\left(\dfrac{1}{\lambda^2}-\rho\right)^2+i \dfrac{|\widetilde{q}|^2}{\lambda^2}    & 2\left(\dfrac{1}{\lambda^3}-\dfrac{\rho}{\lambda}\right)\widetilde{q}-\dfrac{|\widetilde{q}|^2\widetilde{q}}{\lambda}+i \dfrac{\widetilde{q}_{x}}{\lambda}   \vspace{0.1in} \\
        -2\left(\dfrac{1}{\lambda^3}-\dfrac{\rho}{\lambda}\right)\widetilde{q}^{*}+\dfrac{|\widetilde{q}|^2\widetilde{q}^{*}}{\lambda}+i \dfrac{\widetilde{q}^{*}_{x}}{\lambda}
                      &  2 i\left(\dfrac{1}{\lambda^2}-\rho\right)^2-i \dfrac{|\widetilde{q}|^2}{\lambda^2}   \\
                   \end{array}
                   \right). \label{shijian}
\end{eqnarray}
where $\widetilde{q}=\widetilde{q}(x,t)$ is a new potential function and $\widetilde{q}^{*}$ is its complex conjugate.

Therefore, according to the compatibility condition $\widetilde{\varphi}_{xt}=\widetilde{\varphi}_{tx}$ due to Eqs.~(\ref{kongjian}) and (\ref{shijian}) we have
\bes\bee \label{t1}
T_{x}+[T,\, \{U,\, \widetilde{U}\}]=0, \vspace{0.1in} \\
T_{t}+[T,\, \{V,\,  \widetilde{V}\}]=0,
\label{t2}\ene\ees
where we have introduced the generalized Lie bracket $[F, \{G,  \widetilde{G}\}]=FG-\widetilde{G}F$. In particular, the  generalized Lie bracket reduces to the usual Lie bracket $[F, \{G, \widetilde{G}\}]=[F, G]$ for the case $\widetilde{G}=G$.

 Therefore we have
\bee
 \widetilde{U}_t-\widetilde{V}_x+[\widetilde{U},\,\widetilde{V}]=T(U_t-V_x+[U,\, V])T^{-1}=0,
\ene
which yields the same equation (\ref{mnls}) with $q\rightarrow \widetilde{q}$, i.e., $\widetilde{q}$ in the new spectral problem (\ref{kongjian}) and (\ref{shijian}) is a solution of Eq.~(\ref{mnls}).

Hereby, we assume that the Darboux matrix $T(\lambda)$ is of the form
\begin{eqnarray}
 T(\lambda) = \left(
  \begin{array}{cc}  T_{11}(\lambda)    & T_{12}(\lambda) \vspace{0.1in} \\
   T_{21}(\lambda)                     & T_{22}(\lambda) \\
      \end{array}
       \right)= \left(
  \begin{array}{cc}
\lambda^{2N}+\sum\limits_{j=0}^{N-1}A^{(2j)}\lambda^{2j}    &\sum\limits_{j=0}^{N-1}B^{(2j+1)} \lambda^{2j+1}  \vspace{0.1in}\\
-\sum\limits_{j=0}^{N-1}{B^{(2j+1)}}^{*} \lambda^{2j+1}   &\lambda^{2N}+\sum\limits_{j=0}^{N-1}{A^{(2j)}}^{*}\lambda^{2j}
      \end{array}
       \right) . \label{nlsm}
\end{eqnarray}
with the complex functions $A^{(2j)}$ and $B^{(j+1)}\, (j=0,1,...,N-1)$ solving the linear algebraic system $T(\lambda_k)\varphi_k(\lambda_k)=0\, (k=1,2,...,N)$, i.e.,
\bee \label{nlsa}
\left[\lambda_k^{2N}+\sum\limits_{j=0}^{N-1}A^{(2j)}(\lambda_k)\lambda_k^{2j}\right]\varphi_{1}(\lambda_k)+\sum\limits_{j=0}^{N-1}B^{(2j+1)}(\lambda_k) \lambda_k^{2j+1}\varphi_{2}(\lambda_k)=0,
 \vspace{0.1in}\\
\label{nlsb}
-\sum\limits_{j=0}^{N-1}{B^{(2j+1)}}^{*}(\lambda_k) \lambda_k^{2j+1}\varphi_{1}(\lambda_k)+\left[\lambda_k^{2N}+\sum\limits_{j=0}^{N-1}{A^{(2j)}}^{*}(\lambda_k)\lambda_k^{2j}\right] \varphi_{2}(\lambda_k)=0,
\ene
where $\varphi_k(\lambda_k)=(\phi_k(\lambda_k), \psi_k(\lambda_k))^{\rm T}=(\phi_k, \psi_k)^{\rm T}\, (k=1,2,..,N)$ is a solution of the spectral problem (\ref{lax1}) and (\ref{lax2}) for the given spectral parameters $\lambda_k$ and the initial solution $q_0$. The spectral parameters $\lambda_k\, (\lambda_i\neq\lambda_j, i\neq j, i=1,2,...,N)$ are some different parameters suitably chosen such that the determinant of coefficients of Eqs.~(\ref{nlsa}) and (\ref{nlsb}) for the $2N$ variables $A^{(2j)},\, B^{(2j+1)}\, (j=0,1,...,N-1)$ are non-zero. It can be shown that $\pm\lambda_k,\, \pm\lambda^{*}_k\, (k=1,2,...,N)$ are the $4N$ roots of ${\rm det} T(\lambda)=0$ in terms of Eqs.~(\ref{nlsa}) and (\ref{nlsb}), i.e.,
\bee
 {\rm det}\, T(\lambda)=\prod\limits_{k=1}^{N}(\lambda^2-\lambda_k^2)(\lambda^2-\lambda_k^{*2}).
 \ene

The substitution of Eq.~(\ref{nlsm}) into Eqs.~(\ref{t1}) and (\ref{t2}) with the conditions (\ref{nlsa}) and (\ref{nlsb}) yields the following theorem for the $N$-fold Darboux transformation of Eq.~(\ref{mnls}). \\

 {\it Theorem 1.} Let $\varphi_1(\lambda_1),\, \varphi_2(\lambda_2),...,\varphi_N(\lambda_N)$ be $N$ distinct column vector solutions of the corresponding spectral problem (\ref{lax1}) and (\ref{lax2}) for the spectral parameters $\lambda_1,\, \lambda_2,...,\lambda_N$ and the initial solution $q_0(x,t)$ of Eq.~(\ref{mnls}), respectively, then the $N$-fold Darboux transformation of Eq.~(\ref{mnls}) is given by
 \begin{eqnarray} \widetilde{q}_N(x,t)=q_0(x,t)+\dfrac{\partial B^{(2N-1)}}{\partial x}-2i\rho {B^{(2N-1)}}.   \label{dt}
\end{eqnarray}
with $B^{(2N-1)}(x,t)=\dfrac{\Delta B^{(2N-1)}}{\Delta_N},$
\begin{widetext}
\[\Delta_N = \left|\begin{array}{cccccccc}
\lambda_1^{2(N-1)}\phi_1 & {\lambda_1}^{2(N-2)}\phi_1  &\ldots \quad\quad & \phi_1 & {\lambda_1}^{2N-1}\psi_1 &  \lambda_1^{2N-3}\psi_1  &\ldots & \lambda_1\psi_1 \vspace{0.1in} \\
\lambda_2^{2(N-1)}\phi_2 & {\lambda_2}^{2(N-2)}\phi_2  &\ldots \quad\quad & \phi_2 & {\lambda_2}^{2N-1}\psi_2 &  \lambda_2^{2N-3}\psi_2  &\ldots & \lambda_2\psi_2 \vspace{0.1in} \\
\ldots & \ldots      &\ldots  \quad\quad          &\ldots &\ldots & \ldots      &\ldots           &\ldots \vspace{0.1in}  \\
\lambda_N^{2(N-1)}\phi_N & {\lambda_N}^{2(N-2)}\phi_N  &\ldots \quad\quad  & \phi_N & {\lambda_N}^{2N-1}\psi_N & \lambda_N^{2N-3}\psi_N  &\ldots & \lambda_N\psi_N \vspace{0.1in} \\
\lambda_1^{*(2(N-1)}\psi^{*}_1 & {\lambda_1}^{*2(N-2)}\psi_1^{*}  &\ldots \quad\quad  & \psi_1^{*} & -\lambda_1^{*(2N-1)}\phi_1 & -\lambda_1^{*(2N-3)}\phi_1^{*}
 &\ldots & -\lambda_1^{*}\phi_1^{*} \vspace{0.1in}  \\
\lambda_2^{*2(N-1)}\psi^{*}_2 & {\lambda_2}^{*2(N-2)}\psi_2^{*}  &\ldots\quad\quad  & \psi_2^{*} & -\lambda_2^{*(2N-1)}\phi_2 & -\lambda_2^{*(2N-3)}\phi_2^{*}
&\ldots & -\lambda_2^{*}\phi_2^{*} \vspace{0.1in}  \\
\ldots & \ldots      &\ldots \quad\quad           &\ldots &\ldots & \ldots      &\ldots           &\ldots   \vspace{0.1in}  \\
\lambda_N^{*2(N-1)}\psi^{*}_N & {\lambda_N}^{*2(N-2)}\psi_N^{*}  &\ldots \quad\quad  & \psi_N^{*} & -\lambda_N^{*(2N-1)}\phi_N & -\lambda_N^{*(2N-3)}\phi_N^{*}
 &\ldots & -\lambda_N^{*}\phi_N^{*} \vspace{0.1in}  \\
\end{array}\right|,\]
\end{widetext}
and $\Delta B^{(2N-1)}$ is given by the determinant $\Delta_N$ by replacing its $(N+1)$-th column by the column vector $(-\lambda_1^{2N}\phi_1$, $-\lambda_2^{2N}\phi_2$,$\cdots$, $-\lambda_N^{2N}\phi_N$, $-\lambda_1^{*(2N)}\psi_1^{*}$, $-\lambda_2^{*(2N)}\psi_2^{*}$,$\cdots$, $-\lambda_N^{*(2N)}\psi_N^{*})^{\rm T}$.
 \\

Notice that when $\rho=0$, the DT (\ref{dt}) reduces to the DT of the derivative NLS equation, which differs from the known DT~\cite{fan1,kdt}. The $N$-fold Darboux transformation with the initial solution $q_0=0$ (or $q_0$ is an initial plane wave solution) can be used to seek for multi-soliton solutions (or breather solutions) of Eq.~(\ref{mnls}). This is not the main aim of our paper. Our aim is to extend the $N$-fold DT to generate the  generalized perturbation $(n, M)$-fold DT such that multi-rogue wave solutions of Eq.~(\ref{mnls}) are found in terms of determinants.

 In the next sections, we will discuss the generalized perturbation $(n, M)$-fold DT and multi-rogue wave solutions of Eqs.~(\ref{mnls}) and (\ref{gi}) through the Taylor expansion and a limit procedure.

\section{Generalized perturbation $(n, M)$-fold  Darboux transformations and multi-rogue wave solutions}

  In the following we first choose Eq.~(\ref{mnls}) as an example to investigate the `novel' generalized perturbation $(n, M)$-fold DTs in applications of nonlinear integrable equations. In fact, this idea can also be extended to other nonlinear integrable equations and is different from the known ones~\cite{nail09, guo1}.

 Nowadays, to study other types of solutions of Eq.~(\ref{mnls}) such as multi-rogue wave solutions, we need to change some functions $A^{(2j)}$ and $B^{(2j-1)}\, (j=0,1,...,N-1)$ in the above-mentioned Darboux matrix $T$ given by Eq.~(\ref{nlsm}) and initial solution $q_0$ such that we may obtain other types of solutions of Eq.~(\ref{mnls}) in terms of some generalized DTs.

\subsection{Generalized perturbation $(1, N-1)$-fold Darboux transformation method}

Here we still consider the Darboux matrix (\ref{nlsm}), but we only consider one spectral parameter $\lambda=\lambda_1$ not $N\, (N>1)$ distinct spectral parameters $\lambda=\lambda_k\, (k=1,2,...,N)$, in which  the condition $T(\lambda_1)\varphi(\lambda_1)=0$ leads to the linear algebraic system with only two equations
 \bee\label{nlsag}
 \left[\lambda_1^{2N}\!+\!\sum\limits_{j=0}^{N-1}A^{(2j)}\lambda_1^{2j}\right]\!\phi(\lambda_1)
  +\sum\limits_{j=0}^{N-1}B^{(2j+1)}\lambda^{2j+1}_1\psi(\lambda_1)=0,\qquad\quad \\
\label{nlsbg}
 \left[\lambda_1^{*(2N)}\!+\!\sum\limits_{j=0}^{N-1}\!{A^{(2j)}}\lambda_1^{*(2j)}\right]\!\psi^{*}(\lambda_1)\!-\!\sum\limits_{j=0}^{N-1}{B^{(2j+1)}} \lambda_1^{*(2j+1)}\phi^{*}(\lambda_1)\!=\!0,
\ene
where $\varphi(\lambda_1)=(\phi(\lambda_1), \psi(\lambda_1))^{\rm T}$ is a solution of the linear spectral problem (\ref{lax1}) and (\ref{lax2}) with the one spectral parameter $\lambda=\lambda_1$ and the initial solution $q_0(x,t)$ of Eq.~(\ref{mnls}).

For the two linear algebraic equations~(\ref{nlsag}) and (\ref{nlsbg}) containing $2N$ unknown functions $A^{(2j)}$ and $B^{(2j+1)}\, (j=0,1,...,N-1)$, we have two cases for the parameter $N$:
 \begin{itemize}

 \item {} If $N=1$, then we can determine only two complex functions $A^{(0)}$ and $B^{(1)}$ from Eqs.~(\ref{nlsag}) and (\ref{nlsbg}), in which we can not obtain the different functions $A^{(0)}$ and $B^{(1)}$ comparing from the above-mentioned $1$-fold DT such that the `new' solutions can not be found;

 \item {} If $N>1$, then we have $2(N-1)>0$ free functions for $A^{(2j)}$ and $B^{(2j+1)}\, (j=0,1,...,N-1)$. This means that the number of the unknown variables $A^{(2j)}$ and $B^{(2j+1)}$ is larger than one of equations such that we have some free functions, which seems to be useful for the Darboux matrix, but it may be difficult to show the invariant conditions (\ref{t1}) and (\ref{t2}).
 \end{itemize}

For the case $N>1$, we only have two above-given algebraic constraints (\ref{nlsag}) and (\ref{nlsbg}) for $2N$ functions $A^{(2j)}$ and $B^{(2j+1)}\, (j=0,1,...,N-1)$. To determine these $2N$ unknown functions $A^{(2j)}$ and $B^{(2j+1)}$, we need to find additional $2(N-1)$ equations about $2N$ functions $A^{(2j)}$ and $B^{(2j+1)}$ such that we have $2N$ equations about  $2N$ functions $A^{(2j)}$ and $B^{(2j+1)}$, in which we may determine them.
If we can determine these $2N$ functions $A^{(2j)}$ and $B^{(2j+1)}$, then we may obtain `new' solutions of equation (\ref{mnls}) in terms of the Draboux transformation.

Now we start from Eqs.~(\ref{nlsag}) and (\ref{nlsbg}), i.e., $T(\lambda_1)\varphi(\lambda_1)=0$. It follows from Eq.~(\ref{nlsm}) that $T_{ii}\, (i=1,2)$ are two polynomials of degree $2N$ for $\lambda_1$ with the coefficients being $A^{(2j)}$ and $T_{ij}\, (i\not=j,\, i,j=1,2)$ are two polynomials of degree $2N-1$ for $\lambda_1$ with the coefficients being $B^{(2j+1)}$. Since $\varphi(\lambda_1)$ is a column vector solution of the spectral problem (\ref{lax1}) and (\ref{lax2}) for the given spectral parameters $\lambda_k$ and the initial solution $q_0$, thus $\varphi(\lambda_1)$ is a vector function with every component being a function of parameter $\lambda_1$. To generate `new' additional $2(N-1)$ equations
from $T(\lambda_1)\varphi(\lambda_1)=0$, we consider the Taylor expansion of $T(\lambda_1)\varphi(\lambda_1)\big|_{\{\lambda_1\to\lambda_1+\varepsilon\}}=T(\lambda_1+\varepsilon)\varphi(\lambda_1+\varepsilon)$  at $\varepsilon=0$.
We know that \bee \label{nlsp}
 \varphi(\lambda_1+\varepsilon)\!=\!\varphi^{(0)}(\lambda_1)\!+\!\varphi^{(1)}(\lambda_1)\varepsilon
 \!+\!\varphi^{(2)}(\lambda_1)\varepsilon^2\!+\!\cdots,\quad
\ene
where $\varphi^{(k)}(\lambda_1)=\frac{1}{k!}\frac{\partial^k}{\partial \lambda_1^k}\varphi(\lambda_1)=\left(\frac{1}{k!}\frac{\partial^k}{\partial \lambda_1^k}\phi(\lambda_1),\, \frac{1}{k!}\frac{\partial^k}{\partial \lambda_1^k}\psi(\lambda_1)\right)^{\rm T}$ with $\varphi^{(0)}(\lambda_1)=\varphi(\lambda_1)=(\phi(\lambda_1),\, \psi(\lambda_1))^{\rm T}$,\, $(k=0,1,2,...)$,  and
\begin{eqnarray} \label{nlsme}
T(\lambda_1+\varepsilon)=T(\lambda_1)+\sum\limits_{k=1}^{2N} T^{(k)}(\lambda_1) \varepsilon^k=T(\lambda_1)+\sum\limits_{k=1}^{N} T^{(2k-1)} (\lambda_1) \varepsilon^{2k-1}+\sum\limits_{k=1}^{N} T^{(2k)}(\lambda_1) \varepsilon^{2k}, \label{zhan1}
\end{eqnarray}
where $T^{(2k-1)}(\lambda_1)$ and $T^{(2k)}(\lambda_1)$ are given by
\bee
T^{(2k-1)}(\lambda_1)= \left(
  \begin{array}{cc}
C^{2k-1}_{2N} \lambda_1^{2N-2k+1}+\sum\limits_{j=k}^{N-1}C^{2k-1}_{2j} A^{(2j)}\lambda_1^{2j-2k+1}   & \sum\limits_{j=k-1}^{N-1}C^{2k-1}_{2j+1} B^{(2j+1)}\lambda_1^{2j-2k+2}  \vspace{0.1in} \\
-\sum\limits_{j=k-1}^{N-1}C^{2k-1}_{2j+1} {B^{(2j+1)}}^{*}\lambda_1^{2j-2k+2}                   & C^{2k-1}_{2N} \lambda_1^{2N-2k+1}+\sum\limits_{j=k}^{N-1}C^{2k-1}_{2j} {A^{(2j)}}^{*}\lambda_1^{2j-2k+1}\\
      \end{array}
       \right), \ene
\bee
T^{(2k)}(\lambda_1)= \left(
  \begin{array}{cc}
C^{2k}_{2N} \lambda_1^{2N-2k}+\sum\limits_{j=k}^{N-1}C^{2k}_{2j} A^{(2j)}\lambda_1^{2j-2k}   & \sum\limits_{j=k}^{N-1}C^{2k}_{2j+1} B^{(2j+1)}\lambda_1^{2j-2k+1} \vspace{0.1in}\\
-\sum\limits_{j=k}^{N-1}C^{2k}_{2j+1} {B^{(2j+1)}}^{*}\lambda_1^{2j-2k+1}                   & C^{2k}_{2N} \lambda_1^{2N-2k}+\sum\limits_{j=k}^{N-1}C^{2k}_{2j} {A^{(2j)}}^{*}\lambda_1^{2j-2k}\\
      \end{array}
       \right) \ene
 with $C^{k}_j=\dfrac{j (j-1)...(j-k+1)}{k!}$.

Therefore, it follows from Eqs.~(\ref{nlsp}) and (\ref{nlsme}) that we obtain
\bee\label{tp}
T(\lambda_1)\varphi(\lambda_1)\big|_{\{\lambda_1=\lambda_1+\varepsilon\}}= T(\lambda_1+\varepsilon) \varphi(\lambda_1+\varepsilon)
  =\sum_{k=0}^{+\infty}\sum\limits_{j=0}^{k}T^{(j)}(\lambda_1)\varphi^{(k-j)}(\lambda_1)\varepsilon^k. \quad
\ene
Let
 \begin{eqnarray}
\lim\limits_{\varepsilon \to 0}\dfrac{T(\lambda_1+\varepsilon) \varphi(\lambda_1+\varepsilon)}{\varepsilon^s}=
\lim\limits_{\varepsilon \to 0}\dfrac{\sum\limits_{k=0}^{+\infty}\sum\limits_{j=0}^{k}T^{(j)}(\lambda_1)\varphi^{(k-j)}(\lambda_1)\varepsilon^k}
 {\varepsilon^s}=0\,\,\,
(s=0,1, 2,...). \qquad \label{jixian1}
\end{eqnarray}
Since we know that $T(\lambda_1)\varphi(\lambda_1)=T^{(0)}(\lambda_1)\varphi^{(0)}(\lambda_1)=0$ (i.e., Eqs.~(\ref{nlsag}) and (\ref{nlsbg})), thus we have $\sum\limits_{k=0}^{1}\sum\limits_{j=0}^{k}T^{(j)}(\lambda_1)\varphi^{(k-j)}(\lambda_1)=0$ for $s=1$. Similarly, we can also get
$\sum\limits_{k=0}^{s}\sum\limits_{j=0}^{k}T^{(j)}(\lambda_1)\varphi^{(k-j)}(\lambda_1)=0$ for any $s>1$. Since for every $s\in\{0,1,2,...\}$, we can
 have two algebraic equations for unknown functions $A^{(2j)}$ and $B^{(2j+1)}$, thus we choose $s=0,1,...,N-1$
 to generate $2N$ algebraic equations for these $2N$ unknown functions $A^{(2j)}$ and $B^{(2j+1)}$ ($j=0,1,...,N-1$), i.e.,
\bee\label{nls1sys}\begin{array}{r}
T^{(0)}(\lambda_1)\varphi^{(0)}(\lambda_1)=0,  \vspace{0.1in}\\
T^{(0)}(\lambda_1)\varphi^{(1)}(\lambda_1)+T^{(1)}(\lambda_1)\varphi^{(0)}(\lambda_1)=0, \vspace{0.1in} \\
T^{(0)}(\lambda_1)\varphi^{(2)}(\lambda_1)+T^{(1)}(\lambda_1)\varphi^{(1)}(\lambda_1)+T^{(2)}(\lambda_1)\varphi^{(0)}(\lambda_1)=0, \vspace{0.1in} \\
\qquad \cdots\cdots,\qquad\qquad  \vspace{0.1in} \\
\sum\limits_{j=0}^{N-1}T^{(j)}(\lambda_1)\varphi^{(N-1-j)}(\lambda_1)=0,
\end{array}\ene
in which the first vector system, i.e., $T^{(0)}(\lambda_1)\varphi^{(0)}(\lambda_1)=T(\lambda_1)\varphi(\lambda_1)=0$, are just Eqs.~(\ref{nlsag}) and (\ref{nlsbg}), which are required.

Therefore we have obtained the system (\ref{nls1sys}) containing $2N$ algebraic equations  with the $2N$ unknowns functions $A^{(2j)}$ and $B^{(2j+1)}\, (j=0,1,...,N-1)$.
When the eigenvalue $\lambda_1$ is suitably chosen so that the determinant of the coefficients for system~(\ref{nls1sys}) is non-zero, hence the transformation matrix $T$ is uniquely determined by system~(\ref{nls1sys}). It can be shown that Theorem 1 still holds for the Darboux matrix (\ref{nlsm}) with $A^{(2j)},\, B^{(2j+1)} (j=0,1,...,N-1)$ being determined by system (\ref{nls1sys}). Owing to new distinct functions $A^{(2j)}, B^{(2j+1)}$ obtained in the $N$-order Darboux matrix $T$, so we can derive the `new' DT with the same eigenvalue $\lambda=\lambda_1$. Here we call Eqs.~(\ref{dt}) and (\ref{gauge}) associated with new functions $A^{(2j)}, B^{(2j+1)}$ determined by system (\ref{nls1sys}) as a generalized perturbation $(1,N-1)$-fold DT.\\

 {\it Theorem 2.} Let $\varphi(\lambda_1)=(\phi(\lambda_1),\psi(\lambda_1))^{\rm T}$ be a column vector solution of the spectral problem (\ref{lax1}) and (\ref{lax2}) for the spectral parameter $\lambda_1$ and initial solution $q_0(x,t)$ of Eq.~(\ref{mnls}), then the generalized perturbation $(1,N-1)$-fold Darboux transformation of Eq.~(\ref{mnls}) is given by
\bee\label{nls1sol}
 \widetilde{q}_{N}(x,t)=q_0(x,t)+\frac{\partial B^{(2N-1)}}{\partial x}-2i\rho B^{(2N-1)},
\ene
where $B^{(2N-1)}$  is by solving the linear algebraic system ~(\ref{nls1sys}) in terms of the Cramer's rule, i.e., $B^{(2N-1)}=\frac{\Delta^{\epsilon} B^{(2N-1)}}{\Delta_{N}^{\epsilon} }$  with
\bee \label{d1} \Delta_N^{\epsilon}  = \left|\begin{array}{llllllll}
{\lambda}^{2N-2} {\phi^{(0)}} & {\lambda}^{2N-4} {\phi^{(0)}}  &\ldots & {\phi^{(0)}} & {\lambda}^{2N-1} {\psi^{(0)}} & {\lambda}^{2N-3} {\psi^{(0)}}  &\ldots & \lambda{\psi^{(0)}} \vspace{0.1in}\\
\Delta_{2,1}& \Delta_{2,2}  &\ldots &{\phi^{(1)}}  & \Delta_{2,N+1} &  \Delta_{2,N+2} &\ldots &\lambda {\psi^{(1)}}+{\psi^{(0)}}\vspace{0.1in}\\
 \ldots & \ldots      &\ldots           &\ldots &\ldots & \ldots      &\ldots           &\ldots  \vspace{0.1in}\\
\Delta_{N,1}  & \Delta_{N,2} &\ldots &{\phi^{(N-1)}} &\Delta_{N,N+1}  & \Delta_{N,N+2}  &\ldots &\lambda {\psi^{(N-1)}}+{\psi^{(N-2)}} \vspace{0.1in}\\
{\lambda^*}^{(2N-2)} {\psi^{(0)}}^* & {\lambda^*}^{(2N-4)} {\psi^{(0)}}^*  &\ldots & {\psi^{(0)}}^* & -{\lambda^*}^{(2N-1)} {\phi^{(0)}}^* & -{\lambda^*}^{(2N-3)} {\phi^{(0)}}^*  &\ldots & -\lambda^* {\phi^{(0)}}^* \vspace{0.1in}\\
\Delta_{N+2,1} & \Delta_{N+2,2}  &\ldots &{\psi^{(1)}}^*  &\Delta_{N+2,N+1} & \Delta_{N+2,N+2}  &\ldots &-\lambda^* {\phi^{(1)}}^*-{\phi^{(0)}}^*\vspace{0.1in}\\
 \ldots & \ldots      &\ldots           &\ldots &\ldots & \ldots      &\ldots           &\ldots  \vspace{0.1in}\\
\Delta_{2N,1}  & \Delta_{2N,2}  &\ldots &{\psi^{(N-1)}}^* &\Delta_{2N,N+1} & \Delta_{2N,N+2}  &\ldots &-\lambda^* {\phi^{(N-1)}}^*-{\phi^{(N-2)}}^*
\end{array}\right|,\qquad
\ene
\bee\label{d2}
\Delta_{j,s}=\begin{cases}\sum\limits_{k=0}^{j-1}C^{k}_{2N-2s} {\lambda_1}^{(2N-2s-k)} {\phi^{(j-1-k)}}
               \quad {\rm for} \quad 1\leq j,\, s\leq N, \vspace{0.1in} \\
 \sum\limits_{k=0}^{j-1}C^{k}_{4N-2s+1} {\lambda_1}^{(4N-2s+1-k)} {\psi^{(j-1-k)}} \quad {\rm for} \quad 1\leq j\leq N,\, N+1\leq s\leq 2N, \vspace{0.1in}\\
\sum\limits_{k=0}^{j-(N+1)}C^{k}_{2N-2s} {\lambda_1}^{*(2N-2s-k)} {\psi^{(j-N-1-k)*}} \quad {\rm for} \quad
N+1\leq j\leq 2N,\, 1\leq s\leq N, \vspace{0.1in} \\
-\sum\limits_{k=0}^{j-(N+1)}C^{k}_{4N-2s+1} {\lambda_1}^{*(4N-2s+1-k)} {\phi^{(j-N-1-k)*}} \quad {\rm for} \quad
N+1\leq j,\, s \leq 2N \end{cases} \ene
and $\Delta^{\epsilon}  B^{(2N-1)}$ is formed from $\Delta_N^{\epsilon} $ by replacing its
$(N+1)$-th column by the column vector $b=(b_j)_{2N\times 1}$ with
\bee \label{d3}
b_j=\begin{cases} -\sum\limits_{k=0}^{j-1}C^{k}_{2N} {\lambda_1}^{(2N-k)} {\phi^{(j-1-k)}}  \quad {\rm for} \quad  1\leq j\leq N  \vspace{0.1in} \\ -\sum\limits_{k=0}^{j-(N+1)}C^{k}_{2N} {\lambda_1}^{*(2N-k)} {\psi^{(j-N-1-k)*}}  \quad {\rm for} \quad  N+1\leq j\leq 2N.
\end{cases} \ene

Notice that in the name of the generalized perturbation $(1, N-1)$-fold DT, the number $1$ means that we use the number of the distinct spectral parameters and $N-1$ means that the sum of the orders of the highest derivative of the Darboux matrix $T$ in system (\ref{nls1sys})  or the vector eigenfunction $\varphi$ for the used distinct spectral parameters in Eq.~(\ref{jixian1}). \\

\subsection{Generalized perturbation $(n, M)$-fold  Darboux transformation method}

In fact, we can also further extend the above-found generalized perturbation $(1, N-1)$-fold DT, in which we only use one spectral parameter $\lambda=\lambda_1$ and its $m_1$th-order perturbation derivatives of $T(\lambda_1)$ and $\varphi(\lambda_1)$ with $m_1=1,2,...,N-1$. Here we further extend to use $n$ distinct spectral parameters $\lambda_i\, (i=1,2,...,n)$ and their corresponding highest order $m_i\, (m_i=0,1,2,...)$ perturbation derivatives, where these non-negative integers $n,\, m_i$ are required to satisfy $N=n+\sum_{i=1}^nm_i=n+M$ with $M=\sum_{i=1}^nm_i$, where $N$ is the same as one in the Darboux matrix $T$ (\ref{nlsm}).

We consider the Darboux matrix (\ref{nlsm}) and the eigenfunctions $\varphi_i(\lambda_i)\, (i=1,2,...,n)$ are the solutions of the linear spectral problem (\ref{lax1})-(\ref{lax2}) for the spectral parameter $\lambda_i$ and initial solution $q_0$ of Eq.~(\ref{mnls}). Thus we have
\bee\label{kpg}
 T(\lambda_i+\varepsilon) \varphi_i(\lambda_i+\varepsilon)
  \!\!=\!\!\sum_{k=0}^{N-1}\sum\limits_{j=0}^{k}T^{(j)}(\lambda_1)\varphi_i^{(k-j)}(\lambda_i)\varepsilon^k, \quad\,\,
\ene
where $\varphi_i^{(k)}(\lambda_i)=\frac{1}{k!}\frac{\partial^k}{\partial \lambda_i^k}\varphi_i(\lambda)|_{\lambda=\lambda_i}$, and $\varepsilon$ is a small parameter.

It follows from Eq.~(\ref{kpg}) and
 \begin{eqnarray}
\lim\limits_{\varepsilon \to 0}\dfrac{T(\lambda_i+\varepsilon) \varphi_i(\lambda_i+\varepsilon)}{\varepsilon^{k_i}}=0
 \label{jixiankp}
\end{eqnarray}
with $i=1,2,...,n$ and $k_i=0,1,...,m_i$ that we obtain the linear algebraic system with the $2N$ equations ($N=n+\sum_{i=1}^nm_i=n+M$):
\bee\label{kpsysg}\begin{array}{r}
T^{(0)}(\lambda_i)\varphi_i^{(0)}(\lambda_i)=0,  \vspace{0.1in}\\
T^{(0)}(\lambda_i)\varphi_i^{(1)}(\lambda_i)+T^{(1)}(\lambda_i)\varphi_i^{(0)}(\lambda_i)=0, \vspace{0.1in} \\
\qquad \cdots\cdots,\qquad\qquad  \vspace{0.1in} \\
\sum\limits_{j=0}^{m_i}T^{(j)}(\lambda_i)\varphi_i^{(m_i-j)}(\lambda_i)=0,
\end{array}\ene
$i=1,2,...,n$, in which we have some first systems for every index $i$, i.e., $T^{(0)}(\lambda_i)\varphi_i^{(0)}(\lambda_i)=T(\lambda_i)\varphi_i(\lambda_i)=0$ are just some ones in system~(\ref{nls1sys}), but they are different if there exist at least one index $m_i\not=0$. \\

{\it Theorem 3.}  Let $\varphi_i(\lambda_i)=(\phi_i(\lambda_i),\, \psi_i(\lambda_i))^{\rm T}\,\, (i=1,2,...,n)$ be column vector solutions of the spectral problem (\ref{lax1}) and (\ref{lax2}) for the spectral parameters $\lambda_i\, (i=1,2,...,n)$ and the same initial solution $q_0(x,t)$ of Eq.~(\ref{mnls}), respectively, then the generalized perturbation $(n, M)$-fold DT of Eq.~(\ref{mnls}) is given by
\bee\label{kpsolg}
 \widetilde{q}_{N}(x,t)=q_0(x,t)+\frac{\partial B^{(2N-1)}}{\partial x}-2i\rho B^{(2N-1)},
\ene
with $B^{(2N-1)}=\frac{\Delta^{\epsilon(n)}B^{(2N-1)}}{\Delta_N^{\epsilon(n)}}$  being defined by solving the linear algebraic system ~(\ref{kpsysg}) in terms of the Cramer's rule, where $\Delta_N^{\epsilon(n)}={\rm det}([\Delta^{(1)}...\Delta^{(n)}]^{\rm T})$ with
\bee
\Delta^{(i)}=\left[\begin{array}{cccccccc}
{\lambda_i}^{(2N-2)} {\phi_i^{(0)}} & {\lambda_i}^{(2N-4)} {\phi_i^{(0)}}  &\ldots & {\phi_i^{(0)}} & {\lambda_i}^{(2N-1)} {\psi_i^{(0)}} & {\lambda_i}^{(2N-3)} {\psi_i^{(0)}}  &\ldots & \lambda_i\psi_i^{(0)} \vspace{0.1in} \\
\Delta_{2,1}^{(i)}& \Delta_{2,2}^{(i)}  &\ldots &{\phi_i^{(1)}}  & \Delta_{2,N+1}^{(i)} &  \Delta_{2,N+2}^{(i)} &\ldots &
\lambda_i{\psi_i^{(1)}}+\psi_i^{(0)}  \vspace{0.1in}\\
 \vdots & \vdots      &\ddots           &\vdots &\vdots & \vdots      &\ddots           &\vdots  \vspace{0.1in} \\
\Delta_{m_i+1,1}^{(i)}  & \Delta_{m_i+1,2}^{(i)} &\ldots &{\phi_i^{(m_i)}} &\Delta_{m_i+1,N+1}^{(i)}  & \Delta_{m_i+1,N+2}^{(i)}  &\ldots &
\lambda_i\psi_i^{(m_i)}+ \psi_i^{(m_i-1)} \vspace{0.1in}\\
{\lambda_i^*}^{(2N-2)} {\psi_i^{(0)}}^* & {\lambda_i^*}^{(2N-4)} {\psi_i^{(0)}}^*  &\ldots & {\psi_i^{(0)}}^* & -{\lambda_i^*}^{(2N-1)} {\phi_i^{(0)}}^* & -{\lambda_i^*}^{(2N-3)} {\phi_i^{(0)}}^*  &\ldots & -\lambda_i^{*}{\phi_i^{(0)}}^*\vspace{0.1in} \\
\Delta_{m_i+3,1}^{(i)} & \Delta_{m_i+3,2}^{(i)}  &\ldots &{\psi_i^{(1)}}^*  &\Delta_{m_i+3,N+1}^{(i)} & \Delta_{m_i+3,N+2}^{(i)}  &\ldots &-\lambda_i^*{\phi_i^{(1)}}^*-{\phi_i^{(0)}}^* \vspace{0.1in}\\
\vdots & \vdots      &\ddots           &\vdots &\vdots & \vdots      &\ddots           &\vdots  \vspace{0.1in} \\
\Delta_{2(m_i+1),1}^{(i)}  & \Delta_{2(m_i+1),2}^{(i)}  &\ldots &{\psi_i^{(m_i)}}^* &\Delta_{2(m_i+1),N+1}^{(i)} & \Delta_{2(m_i+1),N+2}^{(i)}  &\ldots &-\lambda_i^*{\phi_i^{(m_i)}}^*-{\phi_i^{(m_i-1)}}^* \\
\end{array}\right], \quad \label{mrw1}
\ene
and $\Delta_{j,s}^{(i)}\, (1\leq j\leq 2(m_i+1)$,\, $1\leq s\leq N$ $i=1,2,...,n$) being given by the following formulae:
\bee \label{mrw2}
\Delta_{j,s}^{(i)}=\begin{cases}
 \sum\limits_{k=0}^{j-1}C^{k}_{2N-2s} {\lambda_i}^{2N-2s-k} {\phi_i^{(j-1-k)}}
\quad {\rm for} \quad 1\leq j\leq m_i+1,\, 1\leq s\leq N, \vspace{0.1in} \\
\sum\limits_{k=0}^{j-1}C^{k}_{4N-2s+1} {\lambda_i}^{4N-2s+1-k} {\psi_i^{(j-1-k)}}\quad {\rm for} \quad 1\leq j\leq m_i+1,\, N+1\leq s\leq 2N, \vspace{0.1in}\\
\sum\limits_{k=0}^{j-(N+1)}C^{k}_{2N-2s} {\lambda_i}^{*(2N-2s-k)} {\psi_i^{(j-N-1-k)*}} \quad {\rm for} \quad
m_i+2\leq j\leq 2(m_i+1),\, 1\leq s\leq N, \vspace{0.1in}\\
-\sum\limits_{k=0}^{j-(N+1)}C^{k}_{4N-2s+1} {\lambda_i}^{*(4N-2s+1-k)} {\phi_i^{(j-N-1-k)*}}\quad {\rm for} \quad
m_i+2\leq j\leq 2(m_i+1),\, N+1\leq s\leq 2N \end{cases} \ene
and $\Delta^{\epsilon} {B^{(2N-1)}}$ is formed from the determinant $\Delta_N^{\epsilon(n)}$ by replacing its
$(N+1)$-th column by the column vector $(b^{(1)}\cdots b^{(n)})^{\rm T}$ with $b^{(i)}=(b_j^{(i)})_{2(m_i+1)\times 1}$ and
\bee\label{mrw3}
b_j^{(i)}=\begin{cases}
-\sum\limits_{k=0}^{j-1}C^{k}_{2N} {\lambda_i}^{2N-k} {\phi_i^{(j-1-k)}} \quad {\rm for} \quad  1\leq j\leq m_i+1, \vspace{0.1in}  \\ -\sum\limits_{k=0}^{j-(N+1)}C^{k}_{2N} {\lambda_i}^{*(2N-k)} {\psi_i^{(j-N-1-k)*}} \quad {\rm for} \quad m_i+2\leq j\leq 2(m_i+1).
\end{cases} \ene

Notice that when $n=1$ and $m_1=N-1$, Theorem 3 reduces to the Theorem 2; when $n=N$ and $m_i=0,\,  1\leq i\leq N$, Theorem 3 reduces to  Theorem 1. In the following we will use the generalized perturbation $(1, N-1)$-fold Darboux transformation to investigate multi-rogue wave solutions of the
MNLS equation (\ref{mnls}) from the initial plane wave solution.

\subsection{Multi-rogue wave solutions and parameters controlling}

In this section, we give some multi-rogue wave solutions in terms of determinants, which is different from Ref.~\cite{guo1}, of Eq.~(\ref{mnls}) by means of the generalized perturbation $(1, N-1)$-fold DT. We now consider the `seed' plane wave solution
\bee \nonumber
 q_0(x,t)=ce^{i[ax+(2\rho c^2-ac^2-a^2)t]},
  \ene
where $a$ and $c\not=0$ are real-valued constants,  $a$ is the wave number, and $c$ is the amplitude of the plane wave. It is known that the phase velocity is $(a+c^2-2\rho c^2/a)$, the group velocity is $2a+c^2$, and $|q_0(x,t)|\rightarrow |c|\not=0$ as $|x|, |t|\rightarrow \infty$.

The substitution of the plane wave solution $q_0(x,t)$ into the Lax pair (\ref{lax1}) and (\ref{lax2}) yields their solution for the spectral parameter $\lambda$ as follows:
\begin{eqnarray}
\varphi(\lambda)=\left[
\begin{array}{c}
(C_1 e^{-A}+C_2 e^{A}) e^{B} \vspace{0.1in}\\
i (C_1 e^{A}-C_2 e^{-A})e^{-B}
\end{array}  \label{phi}
\right],
\end{eqnarray}
with
\begin{eqnarray}
\begin{array}{lll}
C_1=C_+,\,\,\, C_2=C_-, \\ \\
C_{\pm}=\dfrac{\sqrt{\pm [2+\lambda^2(a-2\rho)]+\sqrt{\lambda^4(a-2\rho)^2+4\lambda^2(a+c^2-2\rho)+4}}}{\sqrt{2c\lambda}},\\ \\
A=\dfrac{i\sqrt{\lambda^4(a-2\rho)^2+4\lambda^2(a+c^2-2\rho)+4}}{2\lambda^2}\left[x+\left(\dfrac{2}{\lambda^2}-2\rho-c^2-a\right) t+\Theta(\varepsilon)\right], \\
B=-\dfrac{i}{2}[ax+(2\rho c^2-a c^2-a^2)t],\\
 \Theta(\varepsilon)=\sum\limits_{k=1}^{N}(b_k+d_k i)\varepsilon^{2k},
\end{array} \nonumber
\end{eqnarray}
where $b_k,d_k (k=1,2,...,N)$ are real free parameters and $\varepsilon$ is a small parameter.

Nowadays, we fix the spectral parameter $\lambda=\lambda_{1}+\varepsilon^2$ with
\bee\nonumber
\lambda_{1}=\frac{\sqrt{2(a-2\rho)^2\left[2\rho-a-c^2+\sqrt{c^2(2 a-4\rho+c^2)}\right]}}{(a-2\rho)^2},\ene
 in Eq.~(\ref{phi}), then we can expand the vector function $\varphi$ in Eq.~(\ref{phi}) as two Taylor series at $\varepsilon=0$, because the expansion expressions of $\varphi(\varepsilon^2)$ are so complicated. Here we may choose $a=-1,\, \rho=2$, and $c=1$, in which $\lambda_{1}=\frac{3}{5}+\frac{1}{5}i$, to simplify our calculation process. Therefore, we obtain
\begin{eqnarray}\varphi(\varepsilon^2)=\varphi^{(0)}+\varphi^{(1)}\varepsilon^2+\varphi^{(2)}\varepsilon^2
+\varphi^{(3)}\varepsilon^2+\cdots,   \label{e271}
 \end{eqnarray}
where
\begin{eqnarray}
\varphi^{(0)}=\left(
\begin{array}{c}
\phi^{(0)}\\
\psi^{(0)}
\end{array}\right) =\left(
\begin{array}{c}
\sqrt{2}e^{\frac{i}{2}(-x+4t)} \vspace{0.1in}\\
\sqrt{2}e^{-\frac{i}{2}(-x+4t)}
\end{array}  \label{e291}
\right),
\end{eqnarray}

\bee\varphi^{(1)}\!=\!\left[
\begin{array}{c}
\dfrac{3\sqrt{2}}{8}e^{\frac{i}{2}(-x+4t)}[234 t^2-26 x^2-108 x t+12 x+48 t-1+i(156x t- 18 x^2+162t^2- 36t+ 16x-3)] \vspace{0.1in}\\
\dfrac{3\sqrt{2}}{8}e^{-\frac{i}{2}(-x+4t)}[162 t^2-18 x^2+156 x t-16 x+36 t-3+i(26x^2-234t^2+108x t+12x+48t+1)]
\end{array}\right],\,\,\,\,\,
\ene
\bee
\varphi^{(2)}=\left(
\begin{array}{c}
\phi^{(2)}\\
\psi^{(2)}
\end{array}
\right),\qquad
\varphi^{(3)}=\left(
\begin{array}{c}
\phi^{(3)}\\
\psi^{(3)}
\end{array}  \label{phi21}
\right),\, \cdots,
\ene
and $(\phi^{(i)},\psi^{(i)})^{\rm T} (i=2,3)$ are listed in \textbf{Appendix A}.

 By means of Eqs.~(\ref{gauge}), (\ref{dt}), and (\ref{e291}), we can derive new solutions of Eq.~(\ref{mnls}) as follows:
\begin{eqnarray}
\widetilde{q}_N(x,t)=q_0+\dfrac{\partial B^{(2N-1)}}{\partial x}-2i\rho {B^{(2N-1)}}.  \label{solu}
 \end{eqnarray}

It is worth pointing out that we re-derive the seed solution $\widetilde{q}_1(x,t)=-ce^{i[ax+(2\rho c^2-ac^2-a^2)t]}$ for $N=1$. To understand the wave propagation of non-trivial solution (\ref{solu})  of Eq.~(\ref{mnls}) for different parameters, we study their wave structures
as shown in Figs. 1-7 for $N=2,3,4$.

{\it Case I.} \, When $N=2$, according to Theorem 2, we have the first-order rogue wave solution of Eq.~(\ref{mnls})
\begin{eqnarray}
\widetilde{q}_2(x,t)=q_0(x,t)+\dfrac{\partial B^{(3)}}{\partial x}-2i\rho {B^{(3)}}.  \label{rw-10}
 \end{eqnarray}
with $B^{(3)}=\frac{\Delta^{\epsilon}B^{(3)}}{\Delta_2^{\epsilon}}$ and
\[ \Delta_2^{\epsilon}= \left|\begin{array}{cccc}
      \lambda^2 \phi^{(0)} & \phi^{(0)} & \lambda^3 \psi^{(0)}  & \lambda\psi^{(0)} \vspace{0.1in} \\
      \lambda^2\phi^{(1)}+ 2\lambda\phi^{(0)} & \phi^{(1)}  & \lambda^3\psi^{(1)}+ 3\lambda^2\psi^{(0)}  & \lambda \psi^{(1)}+\psi^{(0)} \vspace{0.1in} \\
      {\lambda^*}^2 {\psi^{(0)}}^* & {\psi^{(0)}}^*  &- {\lambda^*}^3 {\phi^{(0)}}^* & -{\lambda^*} {\phi^{(0)}}^* \vspace{0.1in}\\
      {\lambda^*}^2{\psi^{(1)}}^*+ 2{\lambda^*} {\psi^{(0)}}^* & \quad {\psi^{(1)}}^*  & \quad -{\lambda^*}^3{\phi^{(1)}}^*- 3{\lambda^*}^2{\phi^{(0)}}^*  & \quad -{\lambda^*}{\phi^{(1)}}^*-{\phi^{(0)}}^*
\end{array}\right|,\]
\[
\Delta^{\epsilon} B^{(3)}= \left|\begin{array}{cccc}
      \lambda^2 \phi^{(0)} & \phi^{(0)} & -\lambda^4 \phi^{(0)}  & \lambda\psi^{(0)} \vspace{0.1in}\\
      \lambda^2\phi^{(1)}+ 2\lambda\phi^{(0)} & \phi^{(1)}  & -\lambda^4\phi^{(1)}-4\lambda^3\phi^{(0)}  & \lambda \psi^{(1)}+\psi^{(0)} \vspace{0.1in}\\
      {\lambda^*}^2 {\psi^{(0)}}^* & {\psi^{(0)}}^*  &- {\lambda^*}^4 {\psi^{(0)}}^* & -{\lambda^*}{\phi^{(0)}}^* \vspace{0.1in}\\
      {\lambda^*}^2{\psi^{(1)}}^*+ 2{\lambda^*} {\psi^{(0)}}^* & \quad {\psi^{(1)}}^*  & \quad - {\lambda^*}^4{\psi^{(1)}}^*- 4{\lambda^*}^3{\psi^{(0)}}^*  & \quad -{\lambda^*}{\phi^{(1)}}^*-{\phi^{(0)}}^*
\end{array}\right|.
\]

For example, we give the simplification forms of solution (\ref{rw-10}):

{\it Case Ia.} \, For $a=-1,\, c=1,\, \rho=2$,  we have the solution
\bee\label{rw-11}
 \widetilde{q}_{21}(x,t)=\frac{[10(x^2+9t^2)+1-2i(x-t)][10(x^2+9t^2)-3-2i(x+19t)]}{[10(x^2+9t^2)+1+2i(x-t)]^2}e^{i(4t-x)},
\ene

\begin{figure}
	\begin{center}
		{\scalebox{0.38}[0.38]{\includegraphics{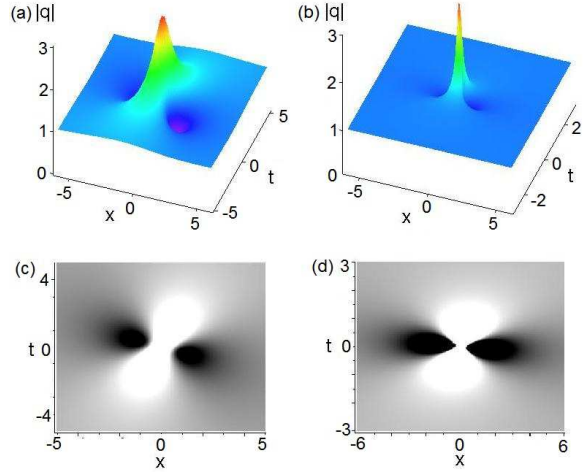}}}
	\end{center}
	\vspace{-0.15in} \caption{\small (color online).  The first-order rogue wave solution $\widetilde{q}_2(x,t)$ given by Eqs.~(\ref{rw-11}) and (\ref{rw-12}) with $a=-1,\, c=1$. (a), (c):  $\rho=0$; (b), (d): $\rho=2$.} \label{fig-rw1}
\end{figure}

{\it Case Ib.} \, For $a= -1, c=1, \rho=0$, we have  the solution
\begin{eqnarray} \begin{array}{ll}  \widetilde{q}_{22}(x,t)=\dfrac{[2(x^2+t^2)+1-2i(x-t)][2(x^2+t^2)-3-2i(x+3t)]}{[2 (x^2+t^2)+1+2i(x-t)]^2}e^{-ix}. \end{array}  \label{rw-12}
\end{eqnarray}
whose wave profiles are exhibited in Fig.~\ref{fig-rw1}.

{\it Case II.} \, When $N=3$,  according to Theorem 2, we have the second-order rogue wave solution of Eq.~(\ref{mnls})
\begin{eqnarray}
\widetilde{q}_3(x,t)=q_0(x,t)+\dfrac{\partial B^{(5)}}{\partial x}-2i\rho {B^{(5)}}.   \label{rw2}
 \end{eqnarray}
with $B^{(5)}=\frac{\Delta^{\epsilon} B^{(5)}}{\Delta_3^{\epsilon}}$ and
\bee\label{delta3} \Delta_3^{\epsilon}= \left|\begin{array}{llllll}
    \lambda^4 \phi^{(0)} &  \lambda^2 \phi^{(0)} & \phi^{(0)} & \lambda^5 \psi^{(0)} & \lambda^3 \psi^{(0)}  & \lambda\psi^{(0)} \vspace{0.1in}\\
  \Delta_{2,1}&   \Delta_{2,2} & \phi^{(1)} & \Delta_{2,4} & \Delta_{2,5}  & \lambda \psi^{(1)}+\psi^{(0)}\vspace{0.1in}\\
  \Delta_{3,1}&    \Delta_{3,2} & \phi^{(2)} & \Delta_{3,4} & \Delta_{3,5}   & \lambda \psi^{(2)}+\psi^{(1)}\vspace{0.1in}\\
  {\lambda^*}^4 {\psi^{(0)}}^* &  {\lambda^*}^2 {\psi^{(0)}}^* & {\psi^{(0)}}^*  &- {\lambda^*}^5 {\phi^{(0)}}^* &- {\lambda^*}^3 {\phi^{(0)}}^* & -{\lambda^*} {\phi^{(0)}}^*\vspace{0.1in}\\
  \Delta_{5,1}  &  \Delta_{5,2}&  {\psi^{(1)}}^*  & \Delta_{5,4} &  \Delta_{5,5} & -{\lambda^*}{\phi^{(1)}}^*-{\phi^{(0)}}^*\vspace{0.1in}\\
  \Delta_{6,1} &     \Delta_{6,2} &  {\psi^{(2)}}^*  &   \Delta_{6,4} &   \Delta_{6,5} & -{\lambda^*}{\phi^{(2)}}^*-{\phi^{(1)}}^*
\end{array}\right|,\ene
\bee\label{32}
\Delta^{\epsilon} B^{(5)}\!=\!\left|\begin{array}{llllll}
    \lambda^4 \phi^{(0)} &  \lambda^2 \phi^{(0)} & \phi^{(0)} & -\lambda^6 \phi^{(0)} &  \lambda^3 \psi^{(0)}  & \lambda\psi^{(0)}\vspace{0.1in}\\
  \Delta_{2,1}&   \Delta_{2,2} & \phi^{(1)} &   -\lambda^6\phi^{(1)}-6\lambda^5\phi^{(0)} & \Delta_{2,5}  & \lambda \psi^{(1)}+\psi^{(0)}\vspace{0.1in}\\
  \Delta_{3,1}&    \Delta_{3,2} & \phi^{(2)} &  -\lambda^6\phi^{(2)}-6\lambda^5\phi^{(1)}-15\lambda^4\phi^{(0)}  & \Delta_{3,5}   & \lambda \psi^{(2)}+\psi^{(1)}\vspace{0.1in}\\
  {\lambda^*}^4 {\psi^{(0)}}^* &  {\lambda^*}^2 {\psi^{(0)}}^* & {\psi^{(0)}}^*  & -{\lambda^*}^6 {\psi^{(0)}}^* &- {\lambda^*}^3 {\phi^{(0)}}^* & -{\lambda^*} {\phi^{(0)}}^*\vspace{0.1in}\\
  \Delta_{5,1}  &  \Delta_{5,2}&  {\psi^{(1)}}^*  & - {\lambda^*}^6{\psi^{(1)}}^*- 6{\lambda^*}^5{\psi^{(0)}}^* &  \Delta_{5,5} & -{\lambda^*}{\phi^{(1)}}^*-{\phi^{(0)}}^*\vspace{0.1in}\\
  \Delta_{6,1} &     \Delta_{6,2} &  {\psi^{(2)}}^*  &    - {\lambda^*}^6{\psi^{(2)}}^*- 6{\lambda^*}^5{\psi^{(1)}}^*- 15{\lambda^*}^4{\psi^{(0)}}^* &   \Delta_{6,5} & -{\lambda^*}{\phi^{(2)}}^*-{\phi^{(1)}}^*
\end{array}\right|, \ene
where
\bee\label{delta33}
\begin{array}{l}\Delta_{2,1}=\lambda^4\phi^{(1)}+4\lambda^3\phi^{(0)},\quad \Delta_{2,2}=\lambda^2\phi^{(1)}+ 2\lambda\phi^{(0)}, \quad \Delta_{2,4}=\lambda^5\psi^{(1)}+ 5\lambda^4\psi^{(0)},  \quad \Delta_{2,5}=\lambda^3\psi^{(1)}+ 3\lambda^2\psi^{(0)}, \vspace{0.1in} \\ \Delta_{3,1}=\lambda^4\phi^{(2)}+4\lambda^3\phi^{(1)}+6\lambda^2\phi^{(0)},\quad \Delta_{3,2}=\lambda^2\phi^{(2)}+ 2\lambda\phi^{(1)}+\phi^{(0)}, \quad \Delta_{3,4}=\lambda^5\psi^{(2)}+ 5\lambda^4\psi^{(1)}+ 10\lambda^3\psi^{(0)}, \vspace{0.1in} \\
 \Delta_{3,5}=\lambda^3\psi^{(2)}+ 3\lambda^2\psi^{(1)}+3\lambda \psi^{(0)},\quad \Delta_{5,1}={\lambda^*}^4{\psi^{(1)}}^*+ 4{\lambda^*}^3 {\psi^{(0)}}^*, \quad \Delta_{5,2}={\lambda^*}^2{\psi^{(1)}}^*+ 2{\lambda^*} {\psi^{(0)}}^*, \vspace{0.1in} \\
  \Delta_{5,4}= - {\lambda^*}^5{\phi^{(1)}}^*- 5{\lambda^*}^4{\phi^{(0)}}^*, \quad
\Delta_{5,5}=- {\lambda^*}^3{\phi^{(1)}}^*- 3{\lambda^*}^2{\phi^{(0)}}^*,  \quad \Delta_{6,1}={\lambda^*}^4{\psi^{(2)}}^*+ 4{\lambda^*}^3 {\psi^{(1)}}^*+6{\lambda^*}^2 {\psi^{(0)}}^*, \vspace{0.1in} \\
 \Delta_{6,2}={\lambda^*}^2{\psi^{(2)}}^*+ 2{\lambda^*} {\psi^{(1)}}^* + {\psi^{(0)}}^*, \quad
 \Delta_{6,4}=- {\lambda^*}^5{\phi^{(2)}}^*- 5{\lambda^*}^4{\phi^{(1)}}^*- 10{\lambda^*}^3{\phi^{(0)}}^*,\vspace{0.1in} \\
  \Delta_{6,5}=- {\lambda^*}^3{\phi^{(2)}}^*- 3{\lambda^*}^2{\phi^{(1)}}^*- 3{\lambda^*}{\phi^{(0)}}^*.
\end{array} \ene

\begin{figure}[!t]
	\begin{center}
	\vspace{0.2in}	{\scalebox{0.38}[0.38]{\includegraphics{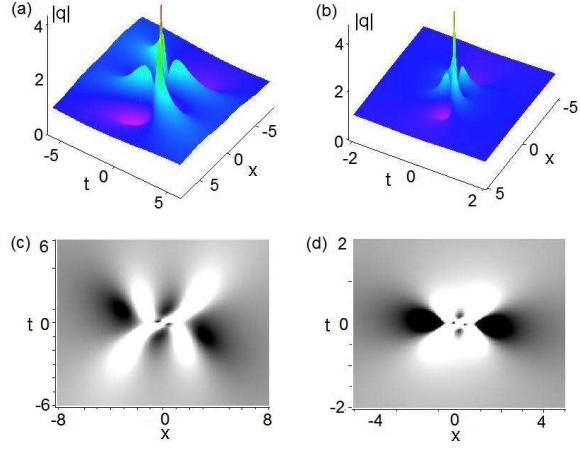}}}
	\end{center}
	\vspace{-0.15in} \caption{\small (color online).  The second-order rogue wave solution given by Eqs.~(\ref{rw21}) and (\ref{rw22}) with $a=-1,\, c=1,\, b_1=d_1=0$. (a), (c):  $\rho=0$; (b), (d): $\rho=2$.} \label{fig-rw2}
\end{figure}

With the aid of symbolic computation, we know that the second-order solution (\ref{rw2}) can explicitly be given from Eqs.~(\ref{rw2}) and (\ref{phi21}) with Eqs.~(\ref{delta3})-(\ref{delta33}), but it is of the long expression about $x,\,t$ and parameters $a,\,c,\, \rho,\, b_1$ and $d_1$. Here we give its explicit expressions for some special parameters:

{\it Case IIa.} \, For the parameters $a=-1,\, c=1,\, \rho=0$ and $b_1=d_1=0$, we have the second-order rogue wave  of Eq.~(\ref{mnls})
\bee\label{rw21}
\widetilde{q}_{31}(x,t)=\frac{G_{11}G_{12}}{(F_{R1}+iF_{I1})^2}e^{-ix},
\ene
with
\bee
F_{R1}=8{x}^{6}+666{t}^{2}-12{x}^{4}-216{x}^{2}{t}^{2}+8{t}^{6}+180{t}^{4}-72xt+24{x}^{4}{t}^{2}+48{x}^{3}t+24\,{x}^{2}{t}^{
4}+48x{t}^{3}+90{x}^{2}+9, \qquad\qquad\qquad \\
F_{I1}=24\,{x}^{5}-24\,{x}^{4}t+48\,{x}^{3}{t}^{2}+54\,x-336\,{t}^{3}-198\,t+48\,{x}^{3}-
24\,{t}^{5}+24\,{t}^{4}x-288\,{t}^{2}x-48\,{x}^{2}{t}^{3}, \qquad\qquad\qquad\quad\qquad\,\,\, \vspace{0.15in} \\
\begin{array}{rl}
G_{11}=& 8x^6-12x^4+180t^4+8t^6+666t^2+90x^2+24x^4t^2+48t^3x+48x^3t-216x^2t^2-72xt+24t^4x^2+9 \vspace{0.1in} \\
&+i(24t^5-48x^3t^2+198t-24x^5+336t^3+48x^2t^3+288xt^2+24x^4t-48x^3-54x-24xt^4), \vspace{0.1in}\\
G_{12}=& -8x^6+60x^4+60t^4-8t^6+486t^2+198x^2-24x^4t^2+144t^3x+144x^3t+504x^2t^2-504xt-24t^4x^2-45 \vspace{0.1in}\\
& +i(24x^5-48x^3+72t^5+528t^3-414t-90x-288x^2t+72x^4t-576xt^2+144x^2t^3+24xt^4+48x^3t^2).
\end{array} \qquad\quad
\ene
The second-order rogue wave solution profile is displayed in Fig.~\ref{fig-rw2}(a) and (c).

\begin{figure}[!t]
	\begin{center}
		{\scalebox{0.38}[0.38]{\includegraphics{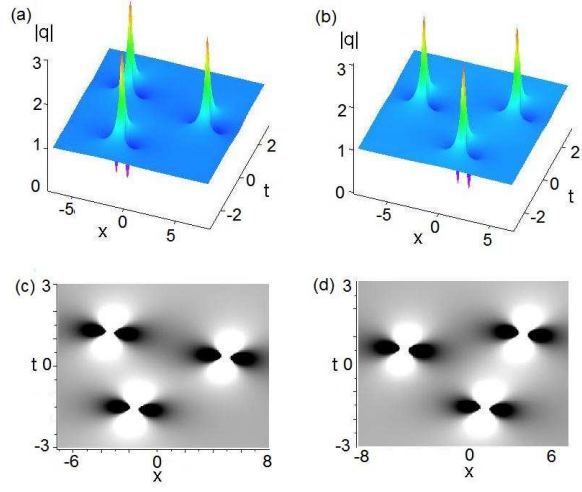}}}
	\end{center}
	\vspace{-0.15in} \caption{\small (color online).  The second-order rogue wave solution $\widetilde{q}_3(x,t)$ given by Eq.~(\ref{rw2}) with $a=-1,\, c=1,\, \rho=2$. (a), (c):  $b_1=-10^3,\, d_1=0$; (b), (d): $b_1=0,\, d_1=10^3$.} \label{fig-rw22}
\end{figure}

{\it Case IIb.} \, For another parameters $a=-1,\, c=1,\, \rho=2$ and $b_1=d_1=0$, we have the second-order rogue wave  of Eq.~(\ref{mnls})
\bee\label{rw22}
\widetilde{q}_{32}(x,t)=\frac{G_{21}G_{22}}{(F_{R2}+iF_{I2})^2}e^{i(4t-x)},
\ene
with \bee
\begin{array}{l}
F_{R2}=306\,{x}^{2}+11106\,{t}^{2}+2160\,x{t}^{3}+240\,{x}^{3}t-19800\,{x}
^{2}{t}^{2}-72\,xt+180\,{x}^{4}+239220\,{t}^{4}+27000\,{x}^{4}{t}^{2} \qquad\qquad \vspace{0.1in}\\
\quad\qquad +729000\,{t}^{6}+1000\,{x}^{6}+243000\,{x}^{2}{t}^{4},\vspace{0.1in} \\
 F_{I2}=600\,{x}^{5}+144\,{x}^{3}-198\,t-10224\,{t}^{3}+54\,x+48600\,{t}^{5}+
288\,{x}^{2}t+48600\,{t}^{4}x-10800\,{x}^{2}{t}^{3}  \vspace{0.1in}\\
\quad\qquad +10800\,{x}^{3}{t}^{2}-4608\,{t}^{2}x-600\,{x}^{4}t,
\end{array} \vspace{0.1in} \\
\begin{array}{rl}
G_{21}=&1000x^6+180x^4+239220t^4+729000t^6+11106t^2+306x^2+27000x^4t^2+2160t^3x+240x^3t \vspace{0.1in}\\
  & -19800x^2t^2-72x+243000t^4x^2+9+i(48600t^5+10224t^3+198t-600x^5-144x^3-54x \vspace{0.1in}\\
  & +10800x^2t^3-48600xt^4-10800x^3t^2-288x^2t+600x^4t+4608xt^2), \vspace{0.1in}\\
G_{22}=&-1000x^6+1020x^4+246780t^4-729000t^6+11358t^2+558x^2-27000x^4t^2+41040t^3x  \vspace{0.1in}\\
 & +4560x^3t+84600x^2t^2-1656xt-243000t^4x^2-45+i(205200x^2t^3+11400x^4t+48600xt^4 \vspace{0.1in}\\
&+10800x^3t^2-17568xt^2-6912x^2t+923400t^5+34416t^3+600x^5-1854t-336x^3-90x).
\end{array} \qquad\qquad\quad
\ene
 The second-order rogue wave solution profile is displayed in Figs.~\ref{fig-rw2}(b) and \ref{fig-rw2}(d).

In fact, the parameters $b_1$ and $d_1$ in solution (\ref{rw2}) can be used to split the second-order rogue wave  (\ref{rw2}) into three first-order rogue waves, whose center points make the triangle exhibited in Fig.~\ref{fig-rw22}. In fact, we find that the sides of this triangle become bigger and bigger as $|b_1|$ and $|d_1|$ increase from zero and the parameter $d_1$ can also control the rotation of the rogue wave profile (see Figs.~\ref{fig-rw22}(b) and \ref{fig-rw22}(d)).

{\it Case III.} \, When $N=4$,  according to Theorem 2, we have the third-order rogue wave solution of Eq.~(\ref{mnls})
\begin{eqnarray}
\widetilde{q}_4(x,t)=q_0(x,t)+\dfrac{\partial B^{(7)}}{\partial x}-2i\rho {B^{(7)}}.   \label{rw3}
\end{eqnarray}
 with $B^{(7)}=\frac{\Delta^{\epsilon} B^{(7)}}{\Delta_5^{\epsilon}}$ and
\bee \label{delta4}
\Delta_5^{\epsilon}= \left|\begin{array}{cccccccc}
    \lambda^6 \phi^{(0)} & \lambda^4 \phi^{(0)} & \lambda^2 \phi^{(0)} & \phi^{(0)} & \lambda^7 \psi^{(0)} & \lambda^5 \psi^{(0)} & \lambda^3 \psi^{(0)}  & \lambda\psi^{(0)}   \vspace{0.1in} \\
    \Delta_{2,1} &\Delta_{2,2} & \Delta_{2,3}& \phi^{(1)}  & \Delta_{2,5} & \Delta_{2,6} &  \Delta_{2,7} & \lambda \psi^{(1)}+ \psi^{(0)}\vspace{0.1in}\\
     \Delta_{3,1} & \Delta_{3,2} &\Delta_{3,3}& \phi^{(2)}  & \Delta_{3,5}& \Delta_{3,6} & \Delta_{3,7}  & \lambda \psi^{(2)}+ \psi^{(1)}\vspace{0.1in}\\
     \Delta_{4,1} & \Delta_{4,2} &\Delta_{4,3} & \phi^{(3)}  & \Delta_{4,5}  & \Delta_{4,6}& \Delta_{4,7} & \lambda \psi^{(3)}+ \psi^{(2)}\vspace{0.1in}\\
    {\lambda^*}^6 {\psi^{(0)}}^* &  {\lambda^*}^4 {\psi^{(0)}}^* &  {\lambda^*}^2 {\psi^{(0)}}^* & {\psi^{(0)}}^*  &-{\lambda^*}^7 {\phi^{(0)}}^* &-{\lambda^*}^5 {\phi^{(0)}}^* &- {\lambda^*}^3 {\phi^{(0)}}^* &- {\lambda^*}{\phi^{(0)}}^*\vspace{0.1in}\\
      \Delta_{6,1} & \Delta_{6,2} & \Delta_{6,3} & {\psi^{(1)}}^*   & \Delta_{6,5} &\Delta_{6,6} & \Delta_{6,7}  & -{\lambda^*} {\phi^{(1)}}^*-{\phi^{(0)}}^*\vspace{0.1in}\\
    \Delta_{7,1} &   \Delta_{7,2}& \Delta_{7,3}  & {\psi^{(2)}}^* & \Delta_{7,5}    &   \Delta_{7,6} & \Delta_{7,7} & -{\lambda^*} {\phi^{(2)}}^*-{\phi^{(1)}}^*\vspace{0.1in}\\
    \Delta_{8,1} &   \Delta_{8,2} & \Delta_{8,3} & {\psi^{(3)}}^* & \Delta_{8,5}   &   \Delta_{8,6}  & \Delta_{8,7}  & -{\lambda^*} {\phi^{(3)}}^*-{\phi^{(2)}}^*
\end{array}\right|,
 \ene
where
 \bee \label{delta42}
 \begin{array}{l}
 \Delta_{2,1}=\lambda^6 \phi^{(1)}+6\lambda^5 \phi^{(0)},\ \Delta_{2,2}=\lambda^4 \phi^{(1)}+4\lambda^3 \phi^{(0)},\ \Delta_{2,3}=\lambda^2 \phi^{(1)}+2\lambda \phi^{(0)},\ \Delta_{2,5}=\lambda^7 \psi^{(1)}+7\lambda^6 \psi^{(0)},\vspace{0.1in}\\
  \Delta_{2,6}=\lambda^5 \psi^{(1)}+5\lambda^4 \psi^{(0)},\ \Delta_{2,7}=\lambda^3 \psi^{(1)}+3\lambda^2 \psi^{(0)},\ \Delta_{3,1}=\lambda^6 \phi^{(2)}+6\lambda^5 \phi^{(1)}+15\lambda^4 \phi^{(0)},\vspace{0.1in}\\
  \Delta_{3,2}=\lambda^4 \phi^{(2)}+4\lambda^3 \phi^{(1)}+6\lambda^2 \phi^{(0)},\ \Delta_{3,3}=\lambda^2 \phi^{(2)}+2\lambda \phi^{(1)}+ \phi^{(0)},\ \Delta_{3,5}=\lambda^7 \psi^{(2)}+7\lambda^6 \psi^{(1)}+21\lambda^5 \psi^{(0)},\vspace{0.1in}\\
   \Delta_{3,6}=\lambda^5 \psi^{(2)}+5\lambda^4 \psi^{(1)}+10\lambda^3 \psi^{(0)},\ \Delta_{3,7}=\lambda^3 \psi^{(2)}+3\lambda^2 \psi^{(1)}+3\lambda \psi^{(0)},\vspace{0.1in}\\
    \Delta_{4,1}=\lambda^6 \phi^{(3)}+6\lambda^5 \phi^{(2)}+15\lambda^4 \phi^{(1)}+20\lambda^3 \phi^{(0)},\ \Delta_{4,2}=\lambda^4 \phi^{(3)}+4\lambda^3 \phi^{(2)}+6\lambda^2 \phi^{(1)}+4\lambda \phi^{(0)},\vspace{0.1in}\\
     \Delta_{4,3}=\lambda^2 \phi^{(3)}+2\lambda  \phi^{(2)}+\phi^{(1)},\ \Delta_{4,5}=\lambda^7 \psi^{(3)}+7\lambda^6 \psi^{(2)}+21\lambda^5 \psi^{(1)}+35\lambda^4 \psi^{(0)},\vspace{0.1in}\\
    \Delta_{4,6}=\lambda^5 \psi^{(3)}+5\lambda^4\psi^{(2)}+10\lambda^3 \psi^{(1)}+10\lambda^2 \psi^{(0)},\ \Delta_{4,7}=\lambda^3 \psi^{(3)}+3\lambda^2\psi^{(2)}+3\lambda \psi^{(1)}+ \psi^{(0)},\vspace{0.1in}\\
    \Delta_{6,1}={\lambda^*}^6 {\psi^{(1)}}^*+6{\lambda^*}^5 {\psi^{(0)}}^*,\ \Delta_{6,2}={\lambda^*}^4{\psi^{(1)}}^*+4{\lambda^*}^3 {\psi^{(0)}}^*,\ \Delta_{6,3}={\lambda^*}^2 {\psi^{(1)}}^*+2{\lambda^*} {\psi^{(0)}}^*,\vspace{0.1in}\\
     \Delta_{6,5}=-{\lambda^*}^7 {\phi^{(1)}}^*-7{\lambda^*}^6 {\phi^{(0)}}^* ,\ \Delta_{6,6}=-{\lambda^*}^5 {\phi^{(1)}}^*-5{\lambda^*}^4 {\phi^{(0)}}^*,\ \Delta_{6,7}=-{\lambda^*}^3 {\phi^{(1)}}^*-3{\lambda^*}^2 {\phi^{(0)}}^*,\vspace{0.1in}\\
     \Delta_{7,1}={\lambda^*}^6 {\psi^{(2)}}^*+6{\lambda^*}^5 {\psi^{(1)}}^*+15{\lambda^*}^4 {\psi^{(0)}}^*,\ \Delta_{7,2}={\lambda^*}^4 {\psi^{(2)}}^*+4{\lambda^*}^3 {\psi^{(1)}}^*+6{\lambda^*}^2{\psi^{(0)}}^* ,\vspace{0.1in}\\
     \Delta_{7,3}={\lambda^*}^2 {\psi^{(2)}}^*+2{\lambda^*} {\psi^{(1)}}^*+{\psi^{(0)}}^* ,\ \Delta_{7,5}=-{\lambda^*}^7 {\phi^{(2)}}^*-7{\lambda^*}^6 {\phi^{(1)}}^*-21{\lambda^*}^5 {\phi^{(0)}}^*,\vspace{0.1in}\\
      \Delta_{7,6}=-{\lambda^*}^5 {\phi^{(2)}}^*-5{\lambda^*}^4 {\phi^{(1)}}^*-10{\lambda^*}^3 {\phi^{(0)}}^*,\ \Delta_{7,7}=-{\lambda^*}^3 {\phi^{(2)}}^*-3{\lambda^*}^2 {\phi^{(1)}}^*-3{\lambda^*} {\phi^{(0)}}^* ,\vspace{0.1in}\\
       \Delta_{8,1}={\lambda^*}^6 {\psi^{(3)}}^*+6{\lambda^*}^5 {\psi^{(2)}}^*+15{\lambda^*}^4 {\psi^{(1)}}^* +20{\lambda^*}^3 {\psi^{(0)}}^*,\ \Delta_{8,2}={\lambda^*}^4 {\psi^{(3)}}^*+4{\lambda^*}^3 {\psi^{(2)}}^*+6{\lambda^*}^2 {\psi^{(1)}}^* +4{\lambda^*} {\psi^{(0)}}^*,\vspace{0.1in}\\
       \Delta_{8,3}={\lambda^*}^2 {\psi^{(3)}}^*+2{\lambda^*}  {\psi^{(2)}}^*+ {\psi^{(1)}}^* ,\ \Delta_{8,5}=-{\lambda^*}^7 {\phi^{(3)}}^*-7{\lambda^*}^6 {\phi^{(2)}}^*-21{\lambda^*}^5 {\phi^{(1)}}^*-35{\lambda^*}^4 {\phi^{(0)}}^*,\vspace{0.1in}\\
       \Delta_{8,6}=-{\lambda^*}^5 {\phi^{(3)}}^*-5{\lambda^*}^4 {\phi^{(2)}}^*-10{\lambda^*}^3 {\phi^{(1)}}^*-10{\lambda^*}^2 {\phi^{(0)}}^*,\ \Delta_{8,7}=-{\lambda^*}^3 {\phi^{(3)}}^*-3{\lambda^*}^2 {\phi^{(2)}}^*-3{\lambda^*} {\phi^{(1)}}^*-{\phi^{(0)}}^*.
       \end{array}
       \ene
Here, $\Delta^{\epsilon} {B^{(7)}}$ is produced from $\Delta_5^{\epsilon}$ by replacing its fifth column with
$(-\lambda^8 \phi^{(0)},-\lambda^8 \phi^{(1)}-8\lambda^7 \phi^{(0)}, -\lambda^8 \phi^{(2)}-8 \lambda^7 \phi^{(1)}-28\lambda^6 \phi^{(0)}, -\lambda^8 \phi^{(3)}-8 \lambda^7 \phi^{(2)}-28 \lambda^6 \phi^{(1)}-56 \lambda^5\phi^{(0)}, -{\lambda^*}^8 {\psi^{(0)}}^*, -{\lambda^*}^8{\psi^{(1)}}^*-8 {\lambda^*}^7{\psi^{(0)}}^*, -{\lambda^*}^8{\psi^{(2)}}^*-8 {\lambda^*}^7 {\psi^{(1)}}^*-28 {\lambda^*}^6{\psi^{(0)}}^*, -{\lambda^*}^8{\psi^{(3)}}^*-8{\lambda^*}^7 {\psi^{(2)}}^*-28 {\lambda^*}^6{\psi^{(1)}}^*-56{\lambda^*}^5 {\psi^{(0)}}^*)^{\rm T}$.

With the aid of symbolic computation, we know that the second-order solution (\ref{rw3}) can explicitly be given from Eqs.~(\ref{rw3}) and (\ref{phi21}) with Eqs.~(\ref{delta4}) and (\ref{delta42}), but it is of the long expression about $x,\,t$ and parameters $a,\,c,\, \rho,\, b_1,\, b_2,\, d_1$ and $d_2$.

\begin{figure}
	\begin{center}
		{\scalebox{0.38}[0.38]{\includegraphics{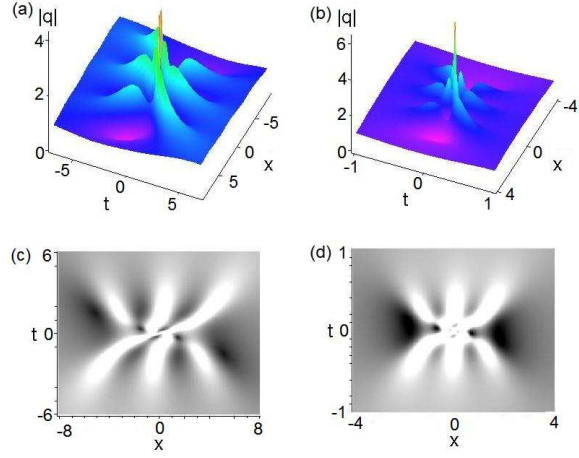}}}
	\end{center}
	\vspace{-0.15in} \caption{\small (color online).  The third-order rogue wave solution $\widetilde{q}_4(x,t)$ given by Eq.~(\ref{rw3}) with $a=-1,\, c=1$ and $b_1=b_2=d_1=d_2=0$. (a), (c):  $\rho=0$; (b), (d): $\rho=2$.} \label{fig-rw3}
\end{figure}

\begin{figure}
	\begin{center}
		{\scalebox{0.38}[0.38]{\includegraphics{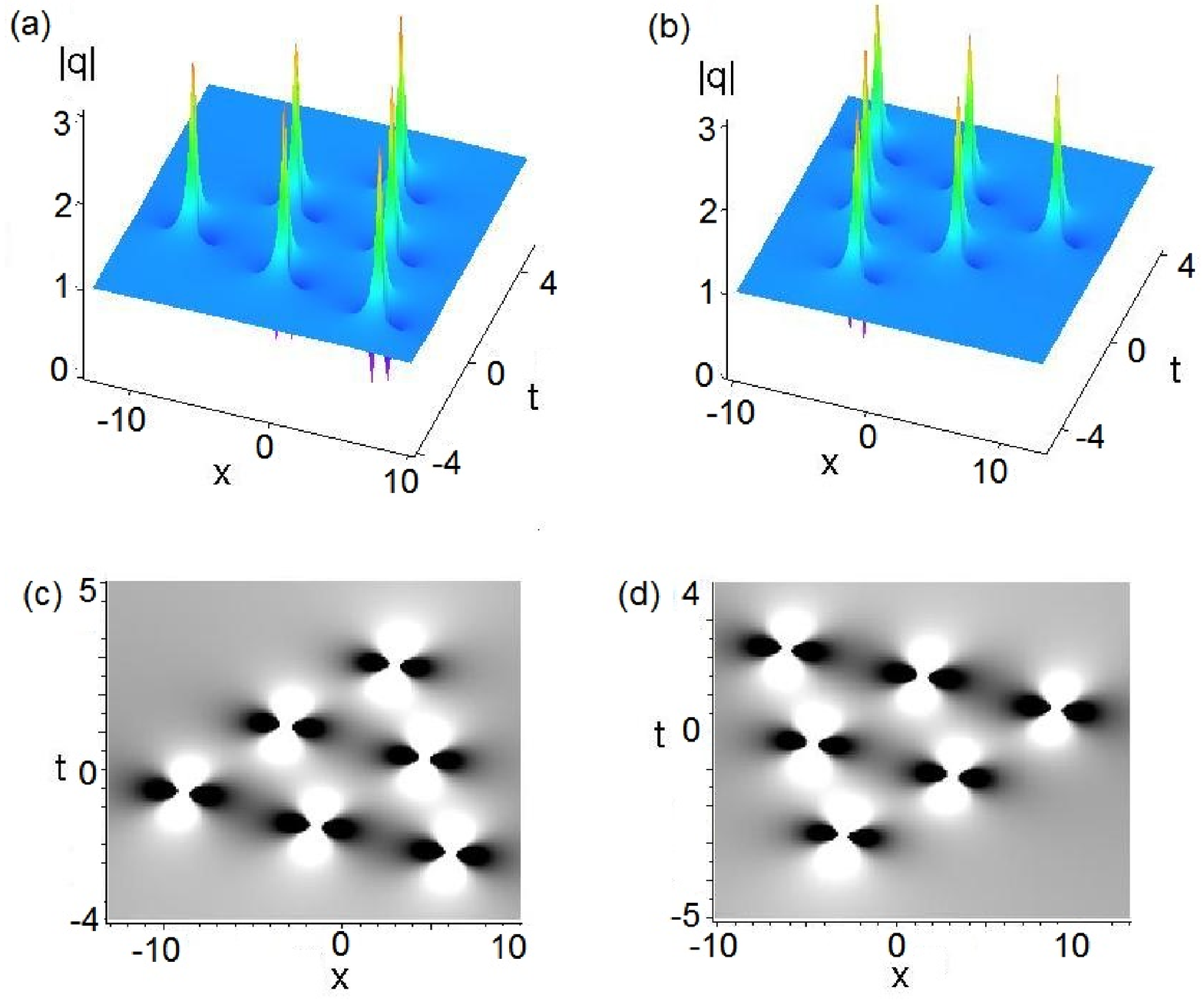}}}
	\end{center}
	\vspace{-0.15in} \caption{\small (color online).  The third-order rogue wave solution $\widetilde{q}_4(x,t)$ given by Eq.~(\ref{rw3}) with $a=-1,\, c=1,\, \rho=2$. (a), (c):  $b_1=10^3,\, d_1=b_2=d_2=0$; (b), (d): $b_1=-10^3,\, d_1=b_2=d_2=0$.} \label{fig-rw32}
\end{figure}

\begin{figure}
	\vspace{0.15in}
	\begin{center}
	{\scalebox{0.38}[0.38]{\includegraphics{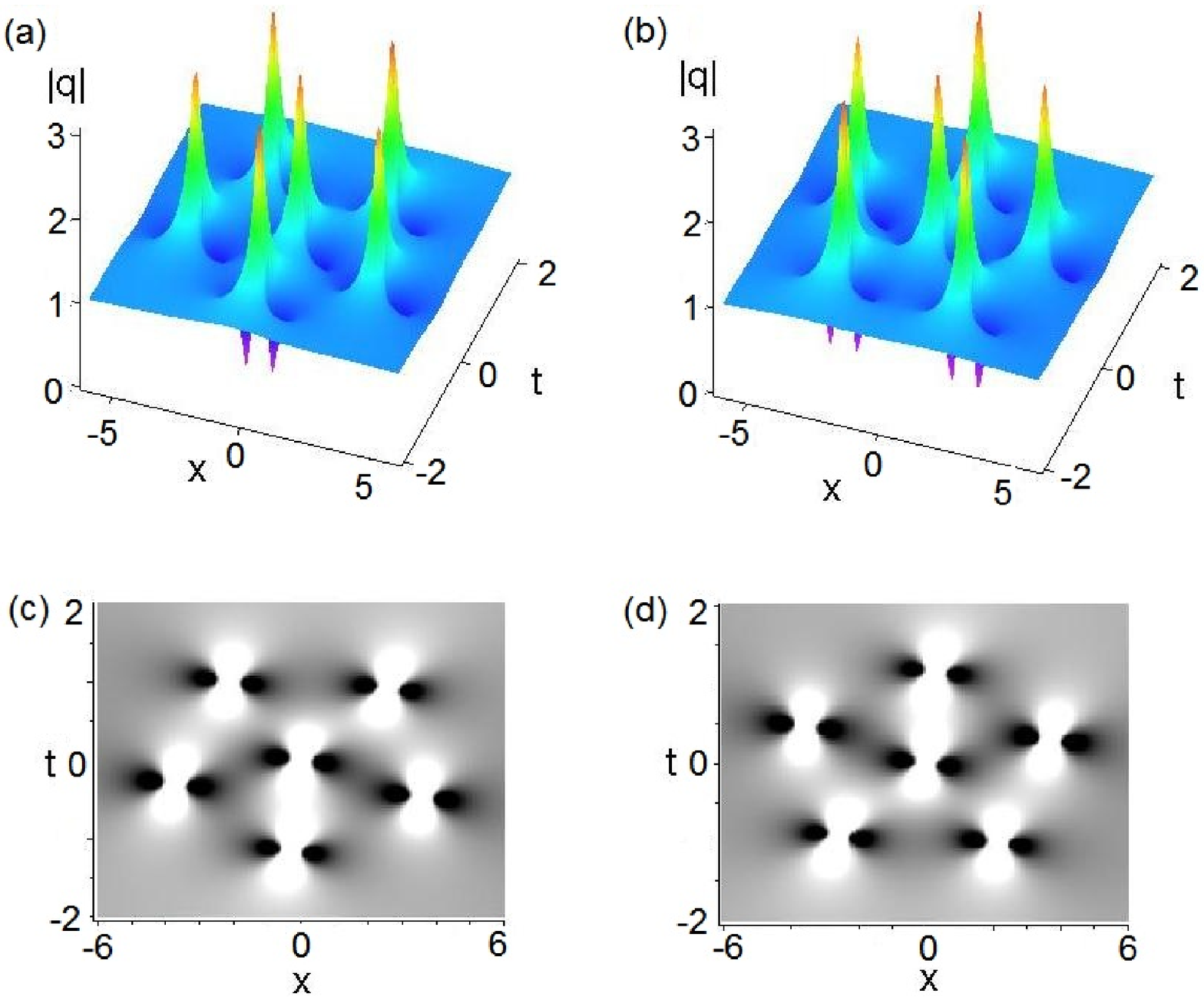}}}
	\end{center}
	\vspace{-0.15in} \caption{\small (color online).  The third-order rogue wave solution $\widetilde{q}_4(x,t)$ given by Eq.~(\ref{rw3}) with $a=-1,\, c=1,\, \rho=2$. (a), (c):  $b_2=10^4,\, d_1=b_1=d_2=0$; (b), (d): $b_2=-10^4,\, d_1=b_1=d_2=0$.} \label{fig-rw33}
\end{figure}

\begin{figure}
	\begin{center}
		{\scalebox{0.54}[0.54]{\includegraphics{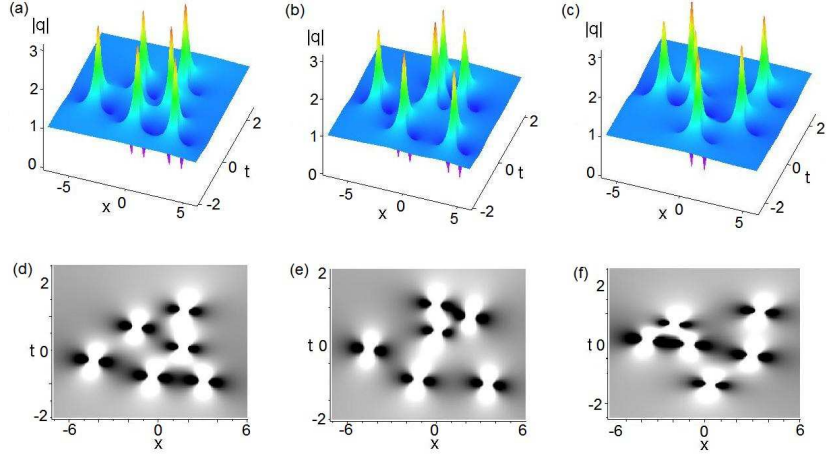}}}
	\end{center}
	\vspace{-0.15in} \caption{\small (color online).  The third-order rogue wave solution $\widetilde{q}_4(x,t)$ given by Eqs.~(\ref{rw3}) and (\ref{rw-12}) with $a=-1,\, c=1,\, \rho=2$. (a), (c): $b_1=100,\, b_2= 3000,\, d_1=d_2=0$; (b), (e): $ b_1=100,\, d_1=0,\, b_2= 10,\, d_2=10^4$,
(c), (f): $b_1=d_2=0,\, d_1=100,\, b_2=10^4$.} \label{fig-rw34}
\end{figure}

For the given parameters $a=-1,\, c=1,\, \rho=0,2$, other parameters $b_1,\,b_2,\, d_1,\,d_2$ can make the third-order rogue wave become the different structures.
\begin{itemize}

  \item{} \, When the parameters $b_1=b_2=d_1=d_2=0$, the the strong interaction of the third-order rogue wave and their corresponding density graphs are shown in Fig.~\ref{fig-rw3}.

\item{} \, When the parameters $b_1=10^3, -10^3,\, d_1=b_2=d_2=0$, the weak interaction of the third-order rogue wave is splitted into six first-order rogue waves, and they array a triangle structure (see Fig.~\ref{fig-rw32}).

\item{} \, When the parameters $b_2=10^4, -10^4,\, d_1=b_2=d_2=0$, the  weak interaction of the third-order rogue wave is also splitted into six first-order rogue waves, but they array a pentagon structure with a first-order rogue wave being almost located in the center of the  pentagon structure (see Fig.~\ref{fig-rw33}).

\item{} \, If we choose one non-zero parameter from two families $\{b_1, d_1\}$ and $\{b_2,\, d_2\}$, respectively, then the third-order rogue wave solution (\ref{rw3}) displays the different structures (see Fig.~\ref{fig-rw34}).
\end{itemize}

\subsection{Dynamical behaviors of multi-rogue wave solutions}

To further illustrate the wave propagations of some above-obtained rogue wave solutions, we here consider the dynamical behaviors of these rogue wave solutions of Eq.~(\ref{mnls}) by comparing these obtained exact multi-rogue wave solutions (e.g., first-order, second-order, and third-order rogue wave solutions) of Eq.~(\ref{mnls}) with their time evolutions using them as initial conditions with or without a small noise via numerical simulations.

\begin{figure}[!ht]
	\begin{center}
	\vspace{0.15in}	{\scalebox{0.75}[0.8]{\includegraphics{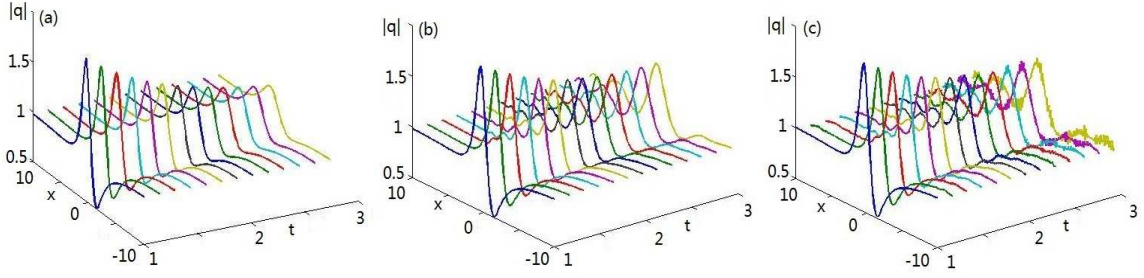}}}
	\end{center}
	\vspace{-0.15in} \caption{\small (color online).  The first-order rogue wave solution $\widetilde{q}_{21}(x,t)$ given by Eq.~(\ref{rw-11}) with $a=-1,\, c=1,\,\rho=0$. (a) exact solution, (b) time evolution using  exact solution (\ref{rw-11}) as the initial condition, (c)
time evolution using  exact solution (\ref{rw-11}) perturbated by a $2\%$ noise as the initial condition.} \label{1-rw-rho0}
\end{figure}

\begin{figure}[!ht]
	\begin{center}
	\vspace{0.15in}	{\scalebox{0.75}[0.8]{\includegraphics{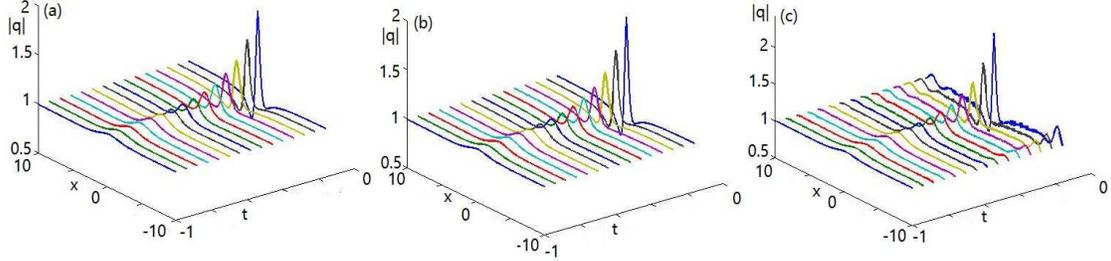}}}
	\end{center}
	\vspace{-0.15in} \caption{\small (color online).  The first-order rogue wave solution $\widetilde{q}_{22}(x,t)$ given by Eq.~(\ref{rw-12}) with $a=-1,\, c=1,\,\rho=2$. (a) exact solution, (b) time evolution of the wave using exact solution (\ref{rw-12}) as the initial condition, (c)
time evolution of the wave using exact solution (\ref{rw-12}) perturbated by a $1\%$ noise as the initial condition.} \label{1-rw-rho2}
\end{figure}

{\it Case 1\, The first-order rogue waves.} \,  For two families of parameters $\{a=-1,\, c=1,\, \rho=0\}$ and  $\{a=-1,\, c=1,\, \rho=2\}$,  Fig.~\ref{1-rw-rho0} and \ref{1-rw-rho2} exhibit the exact first-order rogue wave solution (\ref{rw-11}) of Eq.~(\ref{mnls}),
time evolutions of rogue wave of Eq.~(\ref{mnls}) using exact solution (\ref{rw-11}) and exact solution (\ref{rw-11}) perturbated by a small noise
($2\%$ and $1\%$ for Fig.~\ref{1-rw-rho0}(c) and \ref{1-rw-rho2}(c), respectively) as the initial conditions, respectively.
It follows from Fig.~\ref{1-rw-rho0}(a,b) and \ref{1-rw-rho2}(a,b) that the profiles of time evolutions of rogue waves of Eq.~(\ref{mnls}) without a noise are agree with ones of the corresponding exact rogue wave solutions. Fig.~\ref{1-rw-rho0}(c) displays that the wave profile exhibits the almost  stable propagation, except for some oscillations when time approaches to $3$. Fig.~\ref{1-rw-rho2}(c) illustrates no collapse-instead stable wave propagation, except for some oscillations in the wings of waves when time approaches to $0$.

\begin{figure}[!t]
	\begin{center}
	\vspace{0.15in}	{\scalebox{0.75}[0.8]{\includegraphics{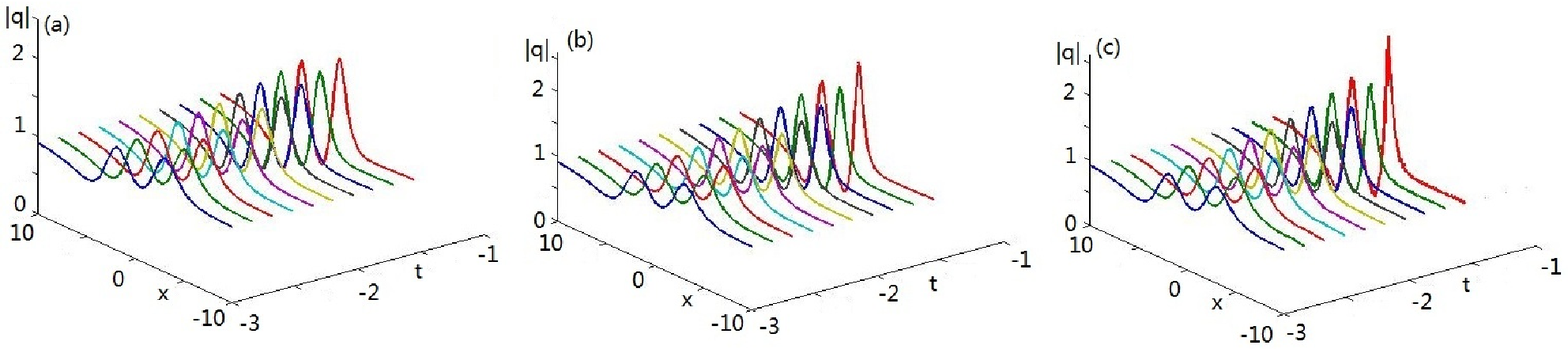}}}
	\end{center}
	\vspace{-0.15in} \caption{\small (color online).  The second-order interactive  rogue wave solution $\widetilde{q}_{31}(x,t)$ given by Eq.~(\ref{rw21}) with $a=-1,\, c=1,\,\rho=0,\, b_1=d_1=0$. (a) exact solution, (b) time evolution of the wave using exact solution (\ref{rw21}) as the initial condition, (c) time evolution of the wave using exact solution (\ref{rw21}) perturbated by a $2\%$ noise as the initial condition.} \label{2-rw-rho0}
\end{figure}

\begin{figure}[!t]
	\begin{center}
	\vspace{0.15in}	{\scalebox{0.75}[0.8]{\includegraphics{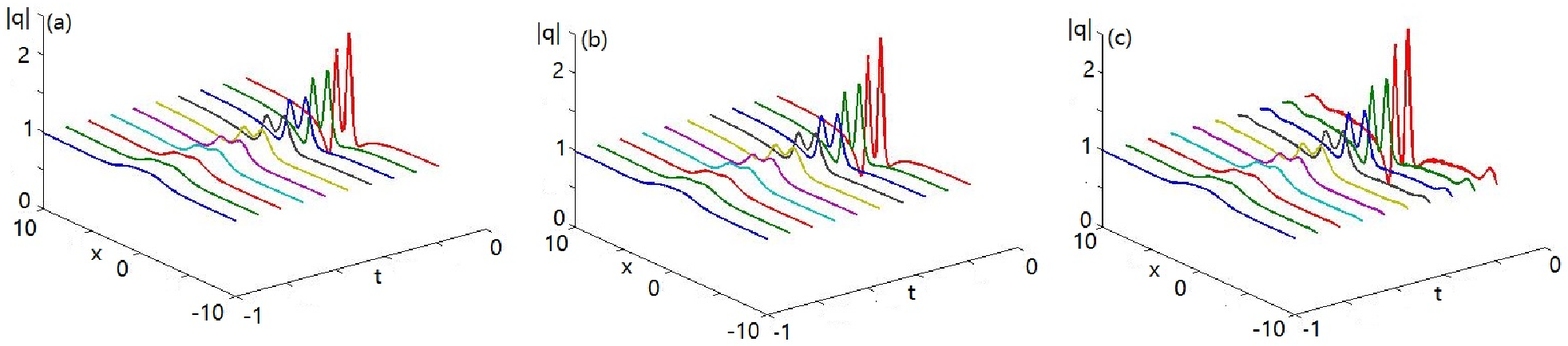}}}
	\end{center}
	\vspace{-0.15in} \caption{\small (color online).  The second-order interactive  rogue wave solution $\widetilde{q}_{32}(x,t)$ given by Eq.~(\ref{rw22}) with $a=-1,\, c=1,\,\rho=2,\, b_1=d_1=0$. (a) exact solution, (b) time evolution of the wave using exact solution (\ref{rw22}) as the initial condition, (c) time evolution of the wave  using exact solution (\ref{rw22}) perturbated by a $1\%$ noise as the initial condition.} \label{2-rw-rho2}
\end{figure}

\begin{figure}[!t]
	\begin{center}
	\vspace{0.15in}	{\scalebox{0.75}[0.8]{\includegraphics{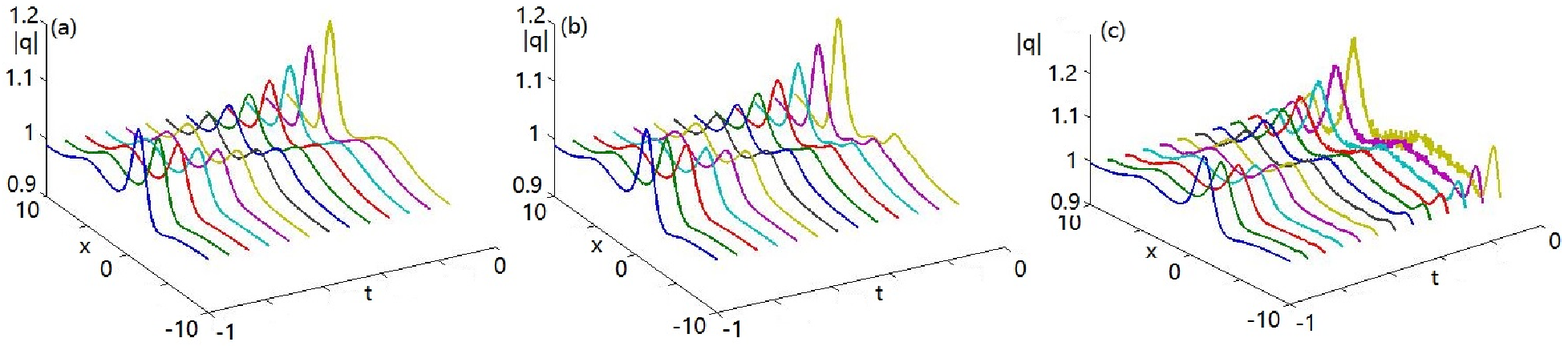}}}
	\end{center}
	\vspace{-0.15in} \caption{\small (color online).  The second-order separatable rogue wave solution $\widetilde{q}_{3}(x,t)$ given by Eq.~(\ref{rw2}) with $a=-1,\, c=1,\,\rho=2,\, b_1=0, d_1=10^3$. (a) exact solution, (b) time evolution of the wave using exact solution (\ref{rw2}) as the initial condition, (c) time evolution of the wave using exact solution (\ref{rw2}) perturbated by a $0.5\%$ noise as the initial condition.} \label{2-rw-rho2-sp}
\end{figure}

{\it Case 2\, The second-order rogue waves.} \, For three families of parameters $\{a=-1,\, c=1,\, \rho=0, \, b_1=d_1=0\}$,  $\{a=-1,\, c=1,\, \rho=2, \, b_1=d_1=0\}$, and $\{a=-1,\, c=1,\, \rho=2, \, b_1=0,\, d_1=10^3\}$,  Figs.~\ref{2-rw-rho0}-\ref{2-rw-rho2-sp}  illustrate the exact second-order rogue wave solution (\ref{rw22}) of Eq.~(\ref{mnls}), time evolution of rogue wave of Eq.~(\ref{mnls}) using exact solution (\ref{rw-12}) and exact solution (\ref{rw22}) perturbated by a small noise (e.g., $2\%$, $1\%$, and $0.5\%$ for Fig.~\ref{2-rw-rho0}(c), ~\ref{2-rw-rho2}(c), and \ref{2-rw-rho2-sp}(c), respectively) as the initial conditions, respectively.  It follows from Fig.~\ref{2-rw-rho0}(a,b), \ref{2-rw-rho2}(a,b), and \ref{2-rw-rho2-sp}(a,b) that the profiles of time evolutions of rogue waves of Eq.~(\ref{mnls}) without a noise are agree with ones of the corresponding exact rogue wave solutions. Fig.~\ref{2-rw-rho0}(c) displays the almost stable wave propagation, however, Figs.~\ref{2-rw-rho2}(c) and \ref{2-rw-rho2-sp}(c)  exhibit the no collapse-instead stable wave propagation, except for some oscillations in the wings of waves when time approaches to $0$.

\begin{figure}[!t]
	\begin{center}
	\vspace{0.15in}	{\scalebox{0.75}[0.8]{\includegraphics{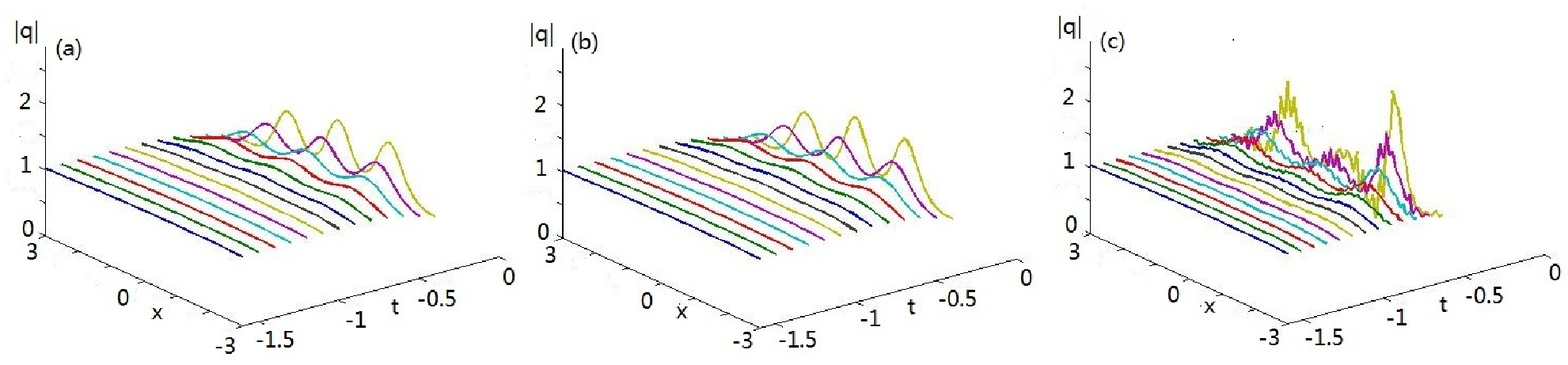}}}
	\end{center}
	\vspace{-0.15in} \caption{\small (color online).  The third-order interactive  rogue wave solution $\widetilde{q}_{4}(x,t)$ given by Eq.~(\ref{rw3}) with $a=-1,\, c=1,\,\rho=2,\, b_1=b_2=d_1=d_2=0$. (a) exact solution, (b) time evolution of the wave using exact solution (\ref{rw3}) as the initial condition, (c) time evolution of the wave using exact solution (\ref{rw3}) perturbated by a $1.5\%$ noise as the initial condition.} \label{3-rw-rho2-b10}
\end{figure}

\begin{figure}[!t]
	\begin{center}
	\vspace{0.15in}	{\scalebox{0.75}[0.8]{\includegraphics{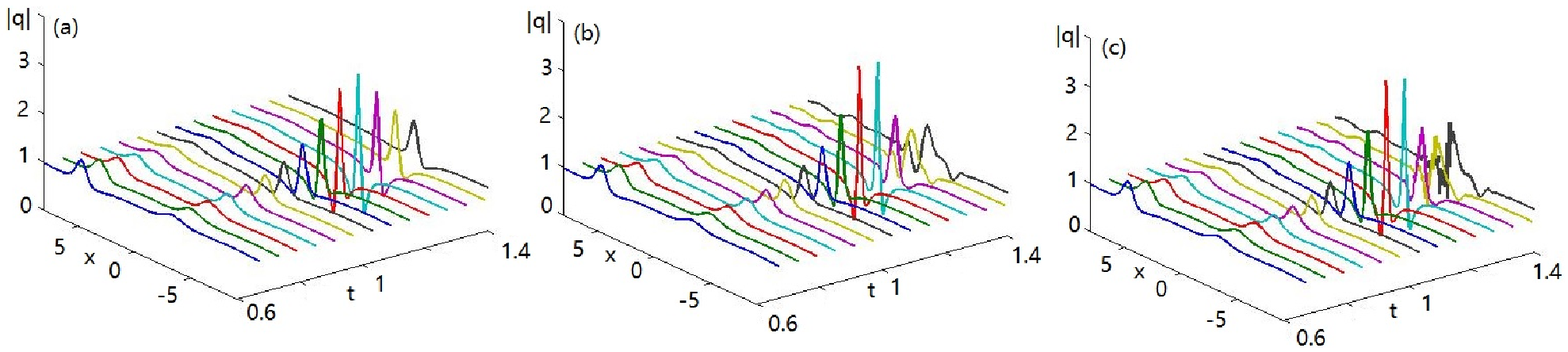}}}
	\end{center}
	\vspace{-0.15in} \caption{\small (color online).  The third-order separatable rogue wave solution $\widetilde{q}_{4}(x,t)$ given by Eq.~(\ref{rw3}) with $a=-1,\, c=1,\,\rho=2,\, b_1=10^3,\, b_2=d_1=d_2=0$. (a) exact solution, (b) time evolution of the wave using exact solution (\ref{rw3}) as the initial condition, (c) time evolution of the wave using exact solution (\ref{rw3}) perturbated by a $0.1\%$ noise as the initial condition.} \label{3-rw-rho2-b1e3}
\end{figure}

\begin{figure}[!t]
	\begin{center}
	\vspace{0.15in}	{\scalebox{0.75}[0.8]{\includegraphics{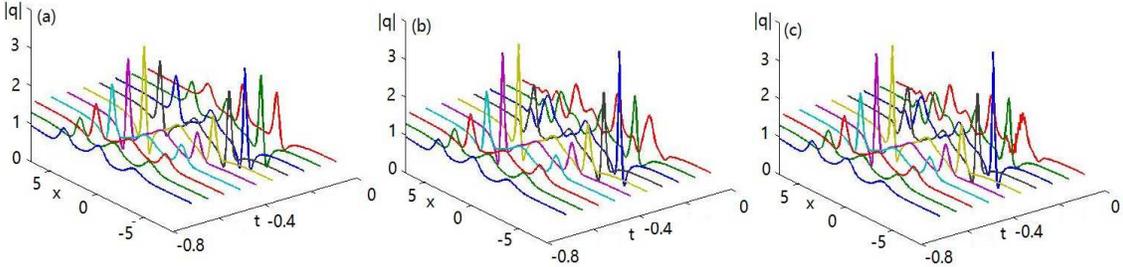}}}
	\end{center}
	\vspace{-0.15in} \caption{\small (color online).  The third-order separatable rogue wave solution $\widetilde{q}_{4}(x,t)$ given by Eq.~(\ref{rw3}) with $a=-1,\, c=1,\,\rho=2,\, b_1=10^4,\, b_2=d_1=d_2=0$. (a) exact solution, (b) time evolution of the wave using exact solution (\ref{rw3}) as the initial condition, (c) time evolution of the wave using exact solution (\ref{rw3}) perturbated by a $0.1\%$ noise as the initial condition.} \label{3-rw-rho2-b1e4}
\end{figure}

{\it Case 3\, The third-order rogue waves.} \, \, For three families of parameters $\{a=-1,\, c=1,\, \rho=2, \, b_1=b_2=d_1=d_2=0\}$,
$\{a=-1,\, c=1,\, \rho=2, \, b_1=10^3,\, b_2=d_1=d_2=0\}$, and $\{a=-1,\, c=1,\, \rho=2, \, b_1=10^4,\, b_2=d_1=d_2=0\}$,
Figs.~\ref{3-rw-rho2-b10}-\ref{3-rw-rho2-b1e4} illustrate the exact third-order rogue wave solution (\ref{rw3}) of Eq.~(\ref{mnls}), time evolution of rogue wave of Eq.~(\ref{mnls}) using exact solution (\ref{rw3}) and exact solution (\ref{rw3}) perturbated by a small noise (e.g., $1.5\%$ for Figs.~\ref{3-rw-rho2-b10}(c) and $0.1\%$ for Figs.~\ref{3-rw-rho2-b1e3}(c) and \ref{3-rw-rho2-b1e4}(c)) as the initial conditions, respectively. It follows from Fig.~\ref{3-rw-rho2-b10}(a,b), \ref{3-rw-rho2-b1e3}(a,b), and \ref{3-rw-rho2-b1e4}(a,b) that the profiles of time evolutions of rogue waves of Eq.~(\ref{mnls}) without a noise are agree with ones of the corresponding exact rogue wave solutions. Fig.~\ref{3-rw-rho2-b10}(c) exhibits the almost unstable wave propagation  from the beginning of about $t=0.5$, however,  Figs.~\ref{3-rw-rho2-b1e3}(c) and \ref{3-rw-rho2-b1e4}(c)  exhibit
 no collapse-instead stable wave propagation, except for some oscillations when time approaches to $1.4$ and $0$, respectively.

\section{The multi-rogue wave solutions of the Gerjikov-Ivanov equation}

\subsection{The generalized perturbation $(n, M)$-fold Darboux transformation method}

The Gerjikov-Ivanov equation (\ref{gi}) is just a zero-curvature equation $U_t-V_x+[U,\,  V]=0$ with $[U,\, V]\equiv UV-VU$ and two $2\times 2$ matrixes $U$ and $V$ satisfying the linear iso-spectral problem (Lax pair)~\cite{xiao}
\begin{eqnarray}
\varphi_x=U \varphi, \qquad U=\left(
                 \begin{array}{cc}
                 -i\lambda^2+\frac{i}{2} |q|^2                          & \lambda q  \vspace{0.1in} \\
                 -\lambda q^{*}                                          & i\lambda^2-\frac{i}{2} |q|^2
                 \end{array}
                 \right),  \qquad\qquad\qquad   \label{lax1g}
\end{eqnarray}
\begin{eqnarray}
 \varphi_{t}=V \varphi, \qquad V=
                 \left(
                 \begin{array}{cc}
        -2 i \lambda^4+i \lambda^2|q|^2+\frac{1}{2}(q q^{*}_x-q^{*} q_x)+\frac{1}{4}|q|^4        & 2 \lambda^3 q+i \lambda q_{x}  \vspace{0.1in}\\
        -2 \lambda^3 q^{*}+i \lambda q^{*}_{x}                   &  2 i \lambda^4-i \lambda^2|q|^2-\frac{1}{2}(q q^{*}_x-q^{*} q_x)-\frac{1}{4}|q|^4  \end{array}
                   \right),  \label{lax2g}
\end{eqnarray}
where $\varphi=(\phi, \psi)^{\rm T}$ is the complex eigenfunction, $\lambda\in \mathbb{C}$ is the spectral parameter, $q=q(x,t)$ denotes the complex potential and is also the solution of Eq.~(\ref{gi}), the subscript denotes the partial derivative with respect to the variables $x,\,t$, and the star stands for the complex conjugate of the corresponding variables.

Similar to the MNLS equation (\ref{mnls}), we choose the same Darboux matrix $T$ given by Eq.~(\ref{nlsm}) to consider the Darboux transformation of Eq.~(\ref{gi}) such that we have the following theorem for the multi-soliton solutions and multi-rogue wave solutions of GI equation (\ref{gi}).\\

{\it Theorem 4.}  Let $\varphi_i(\lambda_i)=(\phi_i(\lambda_i),\, \psi_i(\lambda_i))^{\rm T}\,\, (i=1,2,...,n)$ be column vector solutions of the spectral problem (\ref{lax1g}) and (\ref{lax2g}) for the spectral parameters $\lambda_i\, (i=1,2,...,n)$ and the same initial solution $q_0(x,t)$ of Eq.~(\ref{gi}), respectively, then the generalized perturbation $(n, M)$-fold DT of Eq.~(\ref{gi}) is given by
\bee\label{gis}
 \widetilde{q}_{N}(x,t)=q_0(x,t)+2i B^{(2N-1)},
\ene
where $B^{(2N-1)}=\frac{\Delta^{\epsilon(n)}B^{(2N-1)}}{\Delta_N^{\epsilon(n)}}$ and  $\Delta_N^{\epsilon(n)}={\rm det}([\Delta^{(1)}...\Delta^{(n)}]^{\rm T})$ with $\Delta^{(i)}$ being given by Eq.~(\ref{mrw1}) and $\Delta^{\epsilon} {B^{(2N-1)}}$ is formed from the determinant $\Delta_N^{\epsilon(n)}$ by replacing its $(N+1)$-th column by the column vector $(b^{(1)}\cdots b^{(n)})^{\rm T}$ with $b^{(i)}=(b_j^{(i)})_{2(m_i+1)\times 1}$ and $b_j^{(i)}$ being given by Eq.~(\ref{mrw3}).

\subsection{ The multi-rogue wave solutions}

In the following we give some multi-rogue wave solutions of Eq.~(\ref{gi}) in terms of determinants by use of generalized perturbation $(1, N-1)$-fold DT in Theorem 4. We consider the seed solution of Eq.~(\ref{gi}) in the plane wave form
\bee \label{gip}
 q_0=c e^{i[ax+(\frac{1}{2}c^4-a c^2-a^2)t]},
 \ene
 where $a$ and $c\not=0$ are real-valued constants,  $a$ is the wave number, and $c$ is the amplitude of the plane wave. It is known that the phase velocity is $(a+c^2-c^4/(2a))$, the group velocity is $2a+c^2$, and $|q_0(x,t)|\rightarrow |c|\not=0$ as $|x|, |t|\rightarrow \infty$.

Substituting Eq.~(\ref{gip}) into Eqs.~(\ref{lax1g}) and (\ref{lax2g}), we can give the solution
of Lax pair (\ref{lax1g}) and (\ref{lax2g}) with the spectral parameter $\lambda$ as follows:
\begin{eqnarray}
\varphi(\lambda)=\left[
\begin{array}{c}
i (C_1 e^{-A}-C_2 e^{A})e^{-B} \vspace{0.1in}\\
(C_2 e^{A}+C_1 e^{-A}) e^{B}
\end{array}  \label{e29}
\right],
\end{eqnarray}
with
\begin{eqnarray}
\begin{array}{lll}
C_1=C_{+},\,\, C_2=C_{-},\vspace{0.1in} \\
 C_{\pm}=\dfrac{\sqrt{\pm (2\lambda^2+a-c^2)+\sqrt{(a+2\lambda^2)^2+c^2(c^2-2a)}}}{\sqrt{2c\lambda}}, \vspace{0.1in} \\
 A=i\sqrt{(a+2\lambda^2)^2+c^2(c^2-2a)}\left[\frac{1}{2}x+(\lambda^2-\frac{a}{2}) t+\Theta(\varepsilon)\right], \vspace{0.1in} \\
 B=-\frac{i}{2}[ax+(\frac{1}{2}c^4-a c^2-a^2)t],\vspace{0.1in} \\
 \Theta(\varepsilon)=\sum\limits_{k=1}^{N}(b_k+d_k i) \varepsilon^{2k},
\end{array} \nonumber
\end{eqnarray}
where $b_k,d_k (k=1,2,...,N)$ are real free parameters and $\varepsilon$ is a small parameter.

 Next, we fix
 \bee
  \lambda=\frac{1}{2}\sqrt{2\sqrt{2 c^2 a-c^4}-2a}+\epsilon^2,\ene
for the special case $a=0,\, c=2$, we have $\lambda=1+i+\varepsilon^2$ for simplification, expanding the vector function $\varphi(\varepsilon^2)$ in Eq.~(\ref{e29}) at $\varepsilon=0$, we obtain
\begin{eqnarray}\varphi(\varepsilon^2)=\varphi^{(0)}+\varphi^{(1)}\varepsilon^2+\varphi^{(2)}\varepsilon^4+\varphi^{(3)}\varepsilon^6+\cdots,   \label{e271}
 \end{eqnarray}
where
\begin{eqnarray}
\varphi^{(0)}=\left(
\begin{array}{c}
\phi^{(0)}\\
\psi^{(0)}
\end{array}\right) =\left(
\begin{array}{c}
-\sqrt{2}e^{{4it}} \vspace{0.1in} \\
\sqrt{2}e^{{-4it}}
\end{array}  \label{e30}
\right),
\end{eqnarray}

\bee\varphi^{(1)}\!=\!\left[
\begin{array}{c}
-\dfrac{\sqrt{2}}{8}e^{4it}[32 x^2-512 t^2+256 x t+64 t-1+i(512t^2-32x^2- 16x+ 256x t-1)] \vspace{0.1in}\\
\dfrac{\sqrt{2}}{8}e^{-4it}[32 x^2-512 t^2+256 x t-64 t-1+i(512 t^2- 32 x^2+ 16 x+ 256 x t-1)]
\end{array}\right],\,\,\,\,\,
\ene
\bee
\varphi^{(2)}=\left(
\begin{array}{c}
\phi^{(2)}\\
\psi^{(2)}
\end{array}
\right),\qquad
\varphi^{(3)}=\left(
\begin{array}{c}
\phi^{(3)}\\
\psi^{(3)}
\end{array}  \label{phi2}
\right),\ \ldots
\ene
and $(\phi^{(i)},\psi^{(i)})^{\rm T} (i=2,3)$ are listed in \textbf{Appendix B}.

 For $N=1$ we only deduce the trivial plane wave solution of Eq.~(\ref{gi}). In the following we consider the multi-rogue wave solutions of Eq.~(\ref{gi}) for $N=2,3,4$.

\begin{figure}
	\begin{center}
		{\scalebox{0.38}[0.38]{\includegraphics{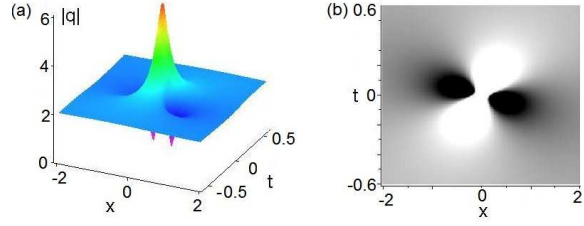}}}
	\end{center}
	\vspace{-0.15in} \caption{\small (color online). (a), (b) The first-order rogue wave solution $\widetilde{q}_2(x,t)$ given by Eq.~(\ref{g-rw1}) with $a=0,\, c=2$.} \label{fig-g-rw1}
\end{figure}

{\it Case I.} \, For $N=2$, we have first-rogue wave solution (\ref{gis}) of Eq.~(\ref{gi}) with $a=0, c=2$:
\begin{eqnarray} \begin{array}{ll}  \widetilde{q}_2(x,t)=2\left[1-\dfrac{4(1+32it)}{1+32(x^2+16t^2)-8i(x-4t)}\right]e^{8it}, \end{array}  \label{g-rw1}
\end{eqnarray}
whose wave profile is shown in Fig.~\ref{fig-g-rw1}. This solution is the same as one in Ref.~\cite{pr14}. \\

\begin{figure}
	\begin{center}
		{\scalebox{0.38}[0.38]{\includegraphics{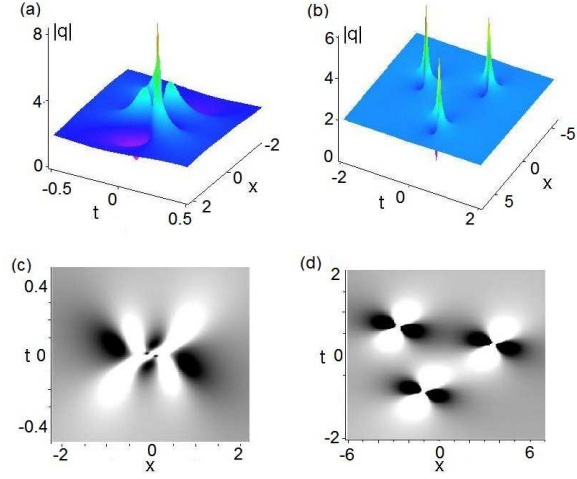}}}
	\end{center}
	\vspace{-0.15in} \caption{\small (color online).  The second-order rogue wave solution $\widetilde{q}_3(x,t)$ with $a=0,\, c=2$.
(a), (c): $b_1=d_1=0$; (b), (d): $ b_1=100,\, d_1=0$.} \label{fig-g-rw2}
\end{figure}

{\it Case II.} \, For $N=3$ we have the second-order rogue wave solutions of Eq.~(\ref{gi}), which is complicated and omitted here. But we give its wave profiles for different parameters. In fact, the parameters $b_1$ and $d_1$ in the second-order rogue wave solution $ \widetilde{q}_3(x,t)$ can be used to split the second-order rogue wave into three first-order rogue waves, whose center points make the triangle exhibited in Figs.~\ref{fig-g-rw2}(b,d). In fact, we find that the sides of this triangle become bigger and bigger as $|b_1|$ and $|d_1|$ increase from zero and the parameter $d_1$ can also control the rotation of the rogue wave profile (see Figs.~\ref{fig-g-rw2}(b,d)).\\

\begin{figure}
	\begin{center}
		{\scalebox{0.5}[0.5]{\includegraphics{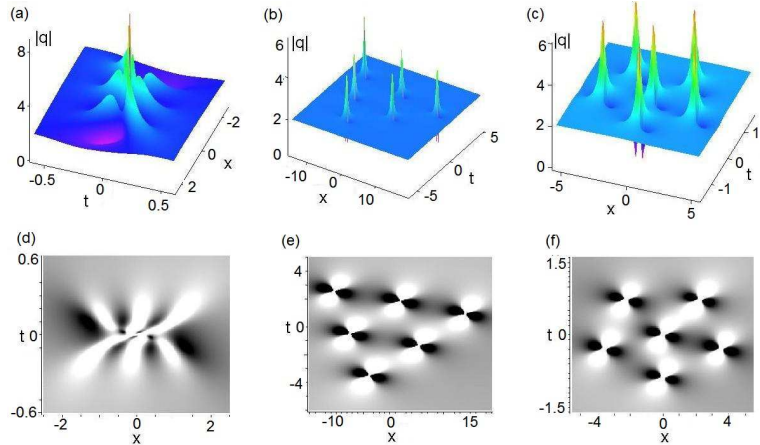}}}
	\end{center}
	\vspace{-0.15in} \caption{\small (color online).  The third-order rogue wave solution $\widetilde{q}_4(x,t)$ with $a=0,\, c=2$.
(a), (d): $b_1=b_2=d_1=d_2=0$; (b), (e): $b_1=1000,\, d_1=b_2=d_2=0$; (c), (f): $b_2=1000,\, b_1=d_1=d_2=0$.} \label{fig-g-rw3}
\end{figure}

{\it Case III.} \, For $N=4$ and the given parameters $a=0,\,c=2$, other parameters $b_1,\,b_2,\, d_1,\,d_2$ can make the third-order rogue wave become the different structures. Fig.~\ref{fig-g-rw3} displays the the interaction of three-order rogue waves.

\begin{itemize}

  \item{} \, When the parameters $b_1=b_2=d_1=d_2=0$, the the interaction of the third-order rogue wave and their corresponding density graphs are shown in Figs.~\ref{fig-g-rw3}(a,d).

\item{} \, When the parameters $b_1=10^3,\, d_1=b_2=d_2=0$, the interaction of the third-order rogue wave is split into six first-order rogue waves, and they array a triangle structure (see Figs.~\ref{fig-g-rw3}(b,e)).

\item{} \, When the parameters $b_2=10^3, \, b_1=d_1=d_2=0$, the interaction of the third-order rogue wave is also split into six first-order rogue waves, but they array a pentagon structure with a first-order rogue wave being almost located in the center of the  pentagon structure (see Figs.~\ref{fig-g-rw3}(c,f)).

\end{itemize}

For other cases $N>4$, we can also obtain the higher-order rogue wave solutions of Eq.~(\ref{gi}), which display the abundant structures. Similar to Eq.~(\ref{mnls}), we can also illustrate the time evolutions of these solutions using numerical simulations, which are omitted here.

\section{Conclusions}

In conclusion, we have presented a novel, simple, and constructive method to find the generalized perturbation $(n, M)$-fold Darboux transformations (DTs) of the modified nonlinear Schr\"odinger equation and the Gerjikov-Ivanov equation in terms of fractional forms of determinants. In particular, we apply the generalized perturbation $(1, N-1)$-fold DTs to find their explicit higher-order rogue wave solutions. The dynamics behaviors of these rogue waves are discussed in detail for the different parameters, which display abundant interesting wave structures including the triangle and pentagon, etc. and may be useful to study the physical mechanism of multi-rogue waves in optics.  Moreover, we study the time evolutions of these obtained multi-rogue wave solutions using numerical simulations.

 For the MNLS equation, if we choose two different spectral parameters
$\lambda_1=(3+i)/5$ and $\lambda_2=2i$, then the higher-order rogue waves can be degraded to lower-order rogue waves. It is still a problem to generate abundant wave structures by choosing more spectral parameters. In fact, the used method can also be extended to seek for multi-rogue wave solutions of other many nonlinear integrable equations such as the NLS equation, KP equation, AB system, AKNS hierarchy, which will be studied in another lecture.

\acknowledgments

The authors would like to thank the referees for their valuable suggestions. This work has been partially supported by the NSFC under Grant Nos. 11375030 and 61178091, the Beijing Natural Science Foundation under Grant No. 1153004, and China Postdoctoral Science Foundation under Grant No. 2015M570161.\\

\noindent\textbf{Appendix A}

$\phi^{(2)}=\frac{1}{64}\sqrt{2}e^{\frac{i}{2}(-x+4t)} (88+ 384 i  b_1-684 x-10620 t+51192 t^3+1944 x t^2+ 57024 i  t^3 x+ 648 i  x^2 t+288 b_1-384 d_1- 1896 i  x^3-72 x^3- 1012 i  x- 2592 i  t d_1- 864 i  x b_1+ 1404 i  x^4+2180 x^2-44964 t^2+ 2340 i  t-28512 x^2 t^2-151632 t^3 x+864 d_1 x+16848 x^3 t+41904 x t-1248 x b_1-3744 t d_1-2592 t b_1+42768 t^4+528 x^4+ 3744 i  t b_1-17064 x^2 t- 21528 i  x t+ 51192 i  x t^2- 1248 i  d_1 x+ 3240 i  x^2- 96552 i  t^2+ 113724 i  t^4+ 288 i  d_1+9 i- 6336 i  x^3 t- 1944 i  t^3- 75816 i  x^2 t^2),$\\

$\psi^{(2)}=-\frac{i}{64}\sqrt{2}e^{-\frac{i}{2}(-x+4t)} (-9-288 i b_1-1012 x+2340 t+528 i x^4-51192 i t^3+88 i-1944 t^3+51192 x t^2-2592 i t b_1+16848 i x^3 t+10620 i t+384 b_1+288 d_1-151632 i t^3 x-1896 x^3-3240 x^2+96552 t^2+75816 x^2 t^2-57024 t^3 x+1248 d_1 x+6336 x^3 t+21528 x t+864 x b_1+2592 t d_1-3744 t b_1-113724 t^4-1404 x^4+648 x^2 t+684 i x-28512 i x^2 t^2-1944 i x t^2+17064 i x^2 t+2180 i x^2+864 i d_1 x-3744 i t d_1-1248 i x b_1+41904 i x t+42768 i t^4+72 i x^3-44964 i t^2+384 i d_1),$\\

$\phi^{(3)}=\frac{1}{2560}\sqrt{2}e^{\frac{i}{2}(-x+4t)} (-2925+1676160 i t d_1+2021760 i x^2 t d_1+100960 x+1709740 t-41636160 t^3+6065280 x t^2 d_1+12862800 x t^2-2280960 i x t^2 d_1+11757312 i t^5+3481660 i x t-27360 b_1+40480 d_1+84480 i d_1 x^3+11520 b_2-15360 d_2-6065280 i x t^2 b_1-572580 i x^4+19852560 i t^3 x-46320 x^3-49920 i d_1 b_1-34560 i x b_2-49920 i d_2 x+2975 i+51840 i x t b_1+64045080 i x^2 t^2+174400 i d_1 x-352170 x^2+33539832 i t^6+5428410 t^2-5770440 x^2 t^2+161215920 t^3 x-259200 d_1 x-10551600 x^3 t-9475380 x t+174400 x b_1+861120 t d_1+1676160 t b_1+21123180 t^4-47060 x^4+7807680 x^2 t-8415576 t^6+24960 d_1^2+11544 x^6-24960 b_1^2-67079664 t^5 x-6065280 t^3 b_1-828144 x^5 t-224640 d_1 x^3+34560 d_1 b_1+14025960 t^4 x^2-1558440 x^4 t^2+34560 d_2 x+24844320 t^3 x^3+84480 x^3 b_1-49920 x b_2-103680 t b_2-149760 t d_2+760320 x^2 t d_1+2021760 x^2 t b_1-2280960 x t^2 b_1-2280960 t^3 d_1-1365120 i x t d_1+259200 i x b_1+15360 i b_2-358800 i x^3 t-145756260 i t^4+530240 i x^3+17280 i d_1^2-6065280 i t^3 d_1-46008 i x^6-17280 i b_1^2+75888 i x^5-461310 i x^2-8640 i x^2 d_1+2280960 i t^3 b_1+6233760 i t^3 x^3+87780 i x+18440784 t^5-48384 x^5-207792 i x^5 t-13063680 i x^2 t^3+725760 i x^4 t-760320 i x^2 t b_1+6211080 i x^4 t^2+149760 i t b_2-18668880 i t^3+2352240 i x^2 t-547680 i t-103680 i t d_2-1365120 x t b_1-8640 x^2 b_1+227520 x^2 d_1-19595520 t^4 x+4354560 x^3 t^2+77760 t^2 b_1-2047680 t^2 d_1-20489760 x^2 t^3+1138320 x^4 t-51840 x t d_1-40480 i b_1+11520 i d_2+30734640 i t^4 x-27360 i d_1-16831152 i t^5 x+31635630 i t^2-32529600 i x t^2+224640 i x^3 b_1-55899720 i t^4 x^2+77760 i t^2 d_1-227520 i x^2 b_1-6829920 i x^3 t^2-861120 i t b_1+2047680 i t^2 b_1),$\\

$\psi^{(3)}=\frac{i}{2560}\sqrt{2}e^{-\frac{i}{2}(-x+4t)} (2975-21123180 i t^4-41636160 i t^3-46320 i x^3-27360 i b_1+11520 i b_2-15360 i d_2-48384 i x^5+18440784 i t^5+40480 i d_1-2021760 i x^2 t b_1-760320 i x^2 t d_1+1709740 i t-51840 i x t d_1-87780 x-5428410 i t^2+100960 i x+352170 i x^2+4354560 i x^3 t^2+47060 i x^4+547680 t+18668880 t^3-2280960 x t^2 d_1+32529600 x t^2+67079664 i t^5 x-24960 i d_1^2-11544 i x^6-14025960 i t^4 x^2+40480 b_1+27360 d_1-15360 b_2-11520 d_2+828144 i x^5 t+24960 i b_1^2-530240 x^3+259200 i d_1 x-34560 i d_2 x+6065280 i t^3 b_1-84480 i x^3 b_1+1558440 i x^4 t^2+7807680 i x^2 t-461310 x^2+31635630 t^2+64045080 x^2 t^2+19852560 t^3 x+174400 d_1 x-358800 x^3 t+3481660 x t+259200 x b_1+1676160 t d_1-861120 t b_1-145756260 t^4-572580 x^4-24844320 i t^3 x^3-2352240 x^2 t-861120 i t d_1+103680 i t b_2+33539832 t^6+17280 d_1^2-46008 x^6-17280 b_1^2-16831152 t^5 x+2280960 t^3 b_1-207792 x^5 t+84480 d_1 x^3-49920 d_1 b_1-55899720 t^4 x^2+6211080 x^4 t^2-49920 d_2 x+6233760 t^3 x^3+224640 x^3 b_1-34560 x b_2+149760 t b_2-103680 t d_2+2021760 x^2 t d_1-760320 x^2 t b_1-6065280 x t^2 b_1-6065280 t^3 d_1+49920 i x b_2-34560 i d_1 b_1+5770440 i x^2 t^2+8415576 i t^6+2925 i-19595520 i t^4 x-2047680 i t^2 d_1+224640 i d_1 x^3-6065280 i x t^2 d_1-161215920 i t^3 x+149760 i t d_2-11757312 t^5-75888 x^5-1676160 i t b_1-1365120 i x t b_1-51840 x t b_1+77760 i t^2 b_1+227520 x^2 b_1+8640 x^2 d_1-30734640 t^4 x+6829920 x^3 t^2-2047680 t^2 b_1-77760 t^2 d_1+13063680 x^2 t^3-725760 x^4 t+1365120 x t d_1-174400 i x b_1+227520 i x^2 d_1-20489760 i x^2 t^3-8640 i x^2 b_1+10551600 i x^3 t+2280960 i x t^2 b_1+1138320 i x^4 t+2280960 i t^3 d_1+12862800 i x t^2+9475380 i x t).$\\

\noindent\textbf{Appendix B}

$\phi^{(2)}= \frac{1}{192}e^{4it}\sqrt{2} (-12+48 x-2112 t-768 d_1+192 x^2+32768 t^3+512 x^3-24576 x t^2-6144 x^2 t-3072 t^2-21504 x t-3072 x b_1-3072 x d_1-12288 t b_1+12288 t d_1-16384 x^3 t+262144 t^3 x+ 262144 i  t^4- 24576 i  x t^2- 98304 i  x^2 t^2+ 1728 i  t+ 768 i  b_1- 12288 i  t d_1- 3072 i  x d_1+ 3072 i  x b_1- 12288 i  t b_1+ 6144 i  x^2 t+ 1152 i  x^2+15 i+ 144 i  x- 67584 i  t^2- 32768 i  t^3+ 512 i  x^3+ 1536 i  x t+ 1024 i  x^4)$,\\

$\psi^{(2)}= -\frac{1}{192}e^{-4it}\sqrt{2} (-12- 768 i  b_1-48 x+2112 t- 1728 i  t+15 i- 67584 i  t^2+768 d_1+192 x^2-32768 t^3-512 x^3+24576 x t^2+6144 x^2 t-3072 t^2-21504 x t-3072 x b_1-3072 x d_1-12288 t b_1+12288 t d_1-16384 x^3 t+262144 t^3 x+ 1024 i  x^4+ 1152 i  x^2+ 32768 i  t^3+ 262144 i  t^4- 144 i  x+ 3072 i  x b_1- 12288 i  t b_1- 3072 i  x d_1- 12288 i  t d_1- 512 i  x^3- 6144 i  x^2 t+ 24576 i  x t^2- 98304 i  x^2 t^2+ 1536 i  x t),$\\

$\phi^{(3)}= -\frac{1}{23040}e^{4it}\sqrt{2} (405- 368640 i  t b_1+ 2949120 i  t d_1 x- 5160960 i  x^3 t+ 1474560 i  t b_2- 125829120 i  t^4 x^2+ 47185920 i  t^2 b_1 x- 2949120 i  t b_1 x+ 11796480 i  x^2 t d_1- 50331648 i  t^5- 92160 i  b_2-6480 x+ 1474560 i  t d_2+ 5160960 i  t d_1- 368640 i  b_1^2- 210240 i  t- 41760 i  x^2- 168960 i  x^4+ 8778240 i  t^2- 169082880 i  t^4- 34560 i  b_1+ 31457280 i  t^3- 15360 i  x^3- 11520 i  d_1+46080 t+ 134217728 i  t^6- 32768 i  x^6- 983040 i  x^4 t+ 368640 i  d_1^2+ 5898240 i  t^2 b_1- 368640 i  x^2 b_1- 368640 i  x^2 d_1+ 5898240 i  t^2 d_1+11796480 x^2 t b_1+ 31457280 i  x^2 t^3+ 145489920 i  t^3 x+ 368640 i  x d_2+ 7864320 i  x^4 t^2- 201326592 i  t^5 x- 786432 i  x^5 t+ 41943040 i  t^3 x^3+ 737280 i  b_1 d_1- 62914560 i  t^3 d_1- 983040 i  x^3 b_1- 92160 i  x d_1- 552960 i  x b_1- 2177280 i  x t- 2949120 i  x^2 t+ 737280 i  x t^2+2949120 t d_1 x+2949120 t b_1 x-47185920 x t^2 d_1- 368640 i  x b_2+ 39813120 i  x^2 t^2-5898240 t^2 d_1+368640 x^2 d_1-11520 b_1+34560 d_1+92160 d_2-64800 x^2-983040 t^3-122880 x^3+17694720 x t^2+184320 x^2 t+10621440 t^2-199680 x^4-176947200 t^4+1624320 x t-92160 x b_1+552960 x d_1+5160960 t b_1+368640 t d_1+4669440 x^3 t-137625600 t^3 x+42762240 x^2 t^2+368640 x b_2+368640 x d_2+1474560 t b_2-1474560 t d_2+737280 b_1 d_1+983040 x^3 d_1-62914560 t^3 b_1+786432 x^5 t+7864320 x^4 t^2-41943040 t^3 x^3-125829120 t^4 x^2+201326592 t^5 x+368640 b_1^2-368640 d_1^2-32768 x^6+134217728 t^6+1395 i+7864320 x^3 t^2-62914560 t^4 x-49152 x^5-368640 x^2 b_1+5898240 t^2 b_1),$\\

$\psi^{(3)}= \frac{1}{23040}e^{-4it}\sqrt{2} (405- 2177280 i  x t+ 145489920 i  t^3 x+6480 x- 737280 i  x t^2-46080 t- 368640 i  x b_2+ 39813120 i  x^2 t^2- 368640 i  b_1^2+11796480 x^2 t b_1- 125829120 i  t^4 x^2-2949120 t d_1 x-2949120 t b_1 x-47185920 x t^2 d_1+5898240 t^2 d_1-368640 x^2 d_1+ 983040 i  x^4 t+11520 b_1-34560 d_1-92160 d_2+ 47185920 i  t^2 b_1 x- 552960 i  x b_1- 368640 i  t b_1-64800 x^2+983040 t^3+122880 x^3-17694720 x t^2-184320 x^2 t+10621440 t^2-199680 x^4-176947200 t^4+1624320 x t-92160 x b_1+552960 x d_1+5160960 t b_1+368640 t d_1+4669440 x^3 t-137625600 t^3 x+42762240 x^2 t^2- 5898240 i  t^2 b_1+368640 x b_2+368640 x d_2+1474560 t b_2-1474560 t d_2+737280 b_1 d_1+983040 x^3 d_1-62914560 t^3 b_1+786432 x^5 t+7864320 x^4 t^2-41943040 t^3 x^3-125829120 t^4 x^2+201326592 t^5 x+368640 b_1^2-368640 d_1^2-32768 x^6+134217728 t^6- 5898240 i  t^2 d_1+ 1474560 i  t d_2+ 1474560 i  t b_2+ 210240 i  t+ 11796480 i  x^2 t d_1- 201326592 i  t^5 x-7864320 x^3 t^2+62914560 t^4 x- 31457280 i  x^2 t^3+49152 x^5- 31457280 i  t^3- 5160960 i  x^3 t+ 50331648 i  t^5+ 737280 i  b_1 d_1+368640 x^2 b_1-5898240 t^2 b_1- 41760 i  x^2+ 2949120 i  x^2 t+ 15360 i  x^3+ 34560 i  b_1+ 5160960 i  t d_1- 168960 i  x^4+ 368640 i  d_1^2- 32768 i  x^6+ 92160 i  b_2- 2949120 i  t d_1 x+ 134217728 i  t^6+ 11520 i  d_1+ 368640 i  x^2 d_1+ 41943040 i  t^3 x^3- 983040 i  x^3 b_1- 62914560 i  t^3 d_1- 169082880 i  t^4+ 2949120 i  t b_1 x+1395 i+ 368640 i  x d_2+ 368640 i  x^2 b_1- 786432 i  x^5 t+ 8778240 i  t^2- 92160 i  x d_1+ 7864320 i  x^4 t^2).$ \\


\begin{thebibliography}{99}

\bibitem{lax} P. D. Lax, Commun. Pure Appl. Math. {\bf XXI}, 467 (1968).

\bibitem{ivs} C. S. Gardner, J. M. Greene, M. D. Kruskal, and R. M. Muria, Phys. Rev. Lett. {\bf 19}, 1095 (1967).

\bibitem{ivs2} M. J. Ablowitz and H. Segur, {\em Solitons and Inverse Scattering Transformation} (SIAM, Philadelphia, 1981).

\bibitem{ivs3} M. J. Ablowitz, P. A. Clarkson, Solitons, {\em Nonlinear Evolution Equations and Inverse Scattering} (Cambridga Univeristy Press, Cambridge, 1991).

\bibitem{RH} P. Deift and E. Trubowitz, Commun. Pure Appl. Math. {\bf 32}, 121 (1979).

\bibitem{dt} V. B. Matveev and M. A. Salle, {\em Darboux Transformation and Solitons} (Springer-Verlag, Berlin, 1991).

\bibitem{dt3} C. H. Gu (ed), {\em Soliton Theory and its Applications} (Srpinger-Verlag, Berlin, 1995) pp122-151.

\bibitem{dt2} C. H. Gu, A. N. Hu, and Z. X. Zhou, {\em Darboux Transformations in Integrable Systems: Theory and their Applications to Geometry} (Springer, Berlin, 2005).

\bibitem{dt82} G. Darboux, C. R. Acd. Sci., Paris {\bf 94}, 1456 (1882).

\bibitem{nail88} N. N. Akhmediev,  V. I. Korneev, N. V. Mitskevich, Zh. Eksp. Teor. Fiz. {\bf 74}, 159 (1988).

\bibitem{nail09} N. Akhmediew, A. Ankiewicz, J. M. Soto-Crespo, Phys. Rev. E {\bf 80}, 026601 (2009).

\bibitem{guo1} B. L. Guo, L. M. Ling, Q. P. Liu, Phys. Rev. E  {\bf 85}, 026607 (2012).


\bibitem{rw1} B. Kibler, J. Fatome, C. Finot, G. Millot, F. Dias, G. Genty, N. Akhmediev, and J.M. Dudley, Nature Phys. {\bf 6}, 1 (2010).
A. Chabchoub, N. Hoffmann, M. Onorato, and N. Akhmediev, Phys. Rev. X {\bf 2}, 011015 (2012).

\bibitem{yang2} Y. Ohta and J. Yang, Proc. Roy. Soc. A. {\bf 468}, 1716 (2012).

\bibitem{yan} Z. Y. Yan, V. V. Konotop, and N. Akhmediev, Phys. Rev. E {\bf 82}, 036610 (2010); Z. Y. Yan, Phys. Lett. A {\bf 374}, 672 (2010);
Z. Y. Yan, Commun. Theor. Phys. {\bf 54}, 947 (2010); Z. Y. Yan, Phys. Lett. A {\bf 375}, 4274 (2011);  Z. Y. Yan, J. Math. Anal. Appl. {\bf 380}, 689 (2011); Z. Y. Yan and C. Dai, J. Opt. {\bf 15}, 064012 (2013); Z. Y. Yan, Nonlinear Dyn. {\bf 79}, 2515 (2015).


\bibitem{rw2} F. Baronio, A. Degasperis, M. Conforti, and S. Wabnitz, Phys. Rev. Lett. {\bf 109}, 044102 (2012);
F. Baronio, M. Conforti, A. Degasperis, S. Lombardo, M. Onorato, and S. Wabnitz, Phys. Rev. Lett. {\bf 113},
034101 (2014).

\bibitem{lak} N. Vishnu Priya, M. Senthilvelan, and M. Lakshmanan, Phys. Rev. E {\bf 89}, 062901 (2014).

\bibitem{ag} G. P. Agrawal, {\em Nonlinear Fiber Optics}, 4th edition (Academic Press, Boston, 2007).

\bibitem{yang} J. K. Yang, {\em Nonlinear waves in integrable and nonintegrable systems} (SIAM, 2010).

\bibitem{kn} D. J. Kaup and A. C. Newell, J. Math. Phys. {\bf 19}, 798 (1978).

\bibitem{gt} D. Mihalache, N. Truta, N.-C. Panoiu, and D.-M. Baboiu, Phys. Rev. A {\bf 47}, 3190 (1993).


\bibitem{mnls1} H. Nakatsuka, D. Grischkowsky and A. C. Balant, Phys. Rev. Lett. {\bf 47}, 910 (1981);
N. Tzoar and M. Jain, Phys. Rev. A {\bf 23}, 1266 (1981).

\bibitem{cp} K. Mio, T. Ogino, K. Minami, and S. Takeda, J. Phys. Soc. Jpn {\bf 41}, 265 (1976).

\bibitem{wm} M. Stiassnie, Wave Motion {\bf 6}, 431 (1984).

\bibitem{pnls} A. I. Maimistov, JETP {\bf 77}, 727 (1993).

\bibitem{wadati} M. Wadati, K. Konno, and Y. H. Ichikawa, J. Phys. Soc. Jpn. {\bf 46}, 1965 (1979).

\bibitem{kon88} V. V. Konotop and  V. E. Vekslerchik, Phys. Lett. A {\bf 131}, 357 (1988).

\bibitem{xiao} Y. Xiao, Commun. Theor. Phys. {\bf 15}, 365 (1991).

\bibitem{huang} Z. Y. Chen and N. N. Huang, Phys. Rev. A {\bf 41}, 4066 (1990).

\bibitem{ns} S. L. Liu and W. Z. Wang, Phys. Rev. E {\bf 48}, 3054 (1993).

\bibitem{liu} J. P. Liu, Commun. Theor. Phys. {\bf 20}, 65 (1993).

\bibitem{ha} Y. Kodama, A. Hasegawa, IEEE J. Quant. Elect. {\bf QE-23}, 510 (1987).

\bibitem{ha2} L. Berg\'e, J. J. Rasmussen, and J. Wyller, J. Phys. A:  Math. Gen. {\bf 29}, 3581 (1996).

\bibitem{ha3} J. S. Hesthaven, {\it et al.,}  J. Phys. A: Math. Gen. {\bf 30}, 8207 (1997).

\bibitem{ha4} L. Berge and S. Skupin, Phys. Rev. E {\bf 71}, 065601R (2005).

\bibitem{guo} B. L. Guo, L. M. Ling, and Q. P. Liu, Stud. Appl. Math. {\bf 130}, 317 (2013).

\bibitem{fan1} E. G. Fan, J. Math. Phys. {\bf 42},  4327 (2001); J. Math. Phys. {\bf 41}, 7769 (2000).

\bibitem{kdt} K. Imai, J. Phys. Soc. Japan {\bf 68}, 355 (1999).

\bibitem{pr14} L. J. Guo, {\it et  al.,} Phys. Scr. {\bf 89}, 035501 (2014).


\end{thebibliography}
\end{document}